\documentclass[useAMS,usenatbib,usegraphicx]{mn2e}
\usepackage{amsmath,amssymb}

\title[Open Cluster Metallicities]
  {Metallicities and Radial Velocities of Five Open Clusters Including a New
  Candidate Member of the Monoceros Stream\thanks{Based on observations
  made with the William Herschel Telescope operated on the island of La
  Palma by the Isaac Newton Group in the Spanish Observatorio del Roque
  de los Muchachos of the Instituto de Astrof\'\i sica de Canarias.}}
\author[Warren \& Cole]
  {Steven R. Warren$^1$ and Andrew A. Cole$^{2,1}$ \\
  $^1$Department of Astronomy, 116 Church Street, S.E., Minneapolis,
  MN 55455; warren@astro.umn.edu\\
  $^2$School of Mathematics \& Physics, University of Tasmania,
  Private Bag 37, Hobart, TAS 7001, Australia; andrew.cole@utas.edu.au}

\begin{document}
\maketitle
\begin{abstract}

{Near infrared spectra of 133 red giant stars from ten Galactic open clusters 
and two Galactic globular clusters spanning 2.2 dex in metallicity and 11 Gyr in age are presented.  
We combine this sample with ten clusters from Cole and collaborators to investigate the Ca II 
triplet line strengths and their
relation to cluster metallicity and position along the red giant branch.
We show that characterizing the stellar surface gravity using $K_{s}$ band
photometry (relative to the horizontal branch) taken from the Two Micron All-Sky Survey allows for
metallicity measurements at least as precise as those derived using
$V$ or $I$ band data.  This has the great advantage that uniform photometry
and reliable astrometry is available for a large number of clusters. Using $K_s$ band photometry
also reduces the effect of 
differential reddening within a given cluster.
We find no significant evidence for age or metallicity effects to the linear Ca II 
triplet - metallicity relationship over the small range in magnitudes studied when homogeneous
reference metallicities are used.  
We derive the first spectroscopic metallicity and new radial velocity 
estimates for five open clusters: Berkeley~81, Berkeley~99, IC~1311, King~2, and NGC~7044.
King~2 has an anomalous radial velocity compared with the local disk population.  We discuss the possibility 
that it is part of the Monoceros tidal stream.
}

\end{abstract}

\begin{keywords}
  techniques: spectroscopic -- stars: abundances -- stars: late-type
-- globular clusters: general -- open clusters:
individual (Berkeley 81, Berkeley 99, IC 1311, King 2, NGC 7044)
\end{keywords}

\section{Introduction}

The metallicities of star clusters and of multi-component stellar populations
in galaxies are among their most important fundamental parameters.  However,
it is often difficult and/or time-consuming to reliably measure the metallicity
or metallicity distribution of a population.  This difficulty is reflected in
the fact that for many open clusters in the Milky Way, the metallicity is still
only known in an extremely vague sense, based on broadband photometry of the
cluster main sequence.  This state of affairs persists even though old open clusters
are widely recognized as testbeds for stellar evolution theory and tracers
of the age-metallicity relation and abundance gradient in the Galaxy
(e.g., \citealt{twa97}; \citealt{friel02}; \citealt{bra06}).  Reliable
spectroscopically-measured abundances, cheaply obtained, are vital to progress
in studies of the kinematics and chemical evolution of clusters and galaxies.

Ideally, spectra of resolution R $\gtrsim$ 30,000
and S/N $\gtrsim$ 100 are used to provide a detailed element-by-element 
assay of the metallicity for multiple stars in each cluster.  However, this
is expensive in both telescope and analysis time, and has only been 
thoroughly carried out for a handful of the thousands of Milky Way clusters
\citep[e.g.,][and references therein]{yon05,car07b,jac07}.  Spectroscopy of the 
near-infrared
calcium triplet (CaT) in K~giants
emerged in the 1990s as a practical alternative
for deriving the overall metal content of distant populations \citep{arm91}, 
empirically calibrated on Galactic globular clusters \citep{rutl97b}.  Recently, this 
technique has begun to be applied to open clusters, both as a path to confirming
its applicability to the younger stars typical of dwarf irregular galaxies and
the metal-rich stars characteristic of the M31 disk and bulge, and as a tool
to explore open cluster metallicities in their own right (e.g., \citealt{cole04};
\citealt{car07}).  A powerful application to the age-metallicity relation and
abundance gradient of the star clusters of the Large Magellanic Cloud was demonstrated
by \citet{gro06}.
The applicability of the CaT technique to composite stellar
populations with a diverse mixture of ages, metallicities, and alpha-element 
enhancements was reconfirmed by \citet{bat08}.

The aim of this paper is twofold:  to increase the flexibility and applicability
of the CaT technique by extending the line-strength--magnitude--metallicity relation
to near-infrared passbands, and to provide the first spectroscopic metallicity and
radial velocity estimates of several distant, understudied open clusters.  The 
advantage of deriving CaT metallicities using $K_s$-band photometry is that surveys
such as the Two Micron All-Sky Survey (2MASS; \citealt{skr06}) make it possible to 
apply the technique to nearly any star accessible to CaT spectroscopy from a 4-metre
class telescope.  In the case of 2MASS, the survey astrometry is also precise enough
to accurately place spectrograph fibres or slitlets, making it a one-stop shop for
lead-up to a multi-object spectroscopy observing campaign. 

We discuss our approach, observations, reductions, and analysis in \S 2.  We show
that the $K_s$ magnitude of the red clump/horizontal branch is at least as precise a
surface gravity proxy as various techniques in the literature that use the $V$ or $I$ 
bands.  The metallicities of seventeen previously studied globular and open clusters
are found to be in good agreement with high-dispersion [Fe/H] values.  Using these
clusters as a calibration sample, we present new abundance and radial velocity 
measurements for five understudied clusters in \S 3.  In most cases the cluster
metallicities are consistent with inferences from colour-magnitude diagrams (CMDs),
and the kinematics and metallicities are suggestive of thick disk membership.
However, the cluster King~2 has a strongly discrepant radial velocity compared
to tracers of the disk; we discuss the possibility that it may be a member of
the Monoceros stream in \S 4.  We conclude with discussion of our results
in \S 5.

\section{Observations and Data Reduction}

The CaT has risen to a pre-eminent position among 
metallicity indicators for faint red giants.  The large-scale 
study of globular clusters in \citet{rutl97a} and \citet{rutl97b}
led to the work of \citet{cole04}, \citet{car07}, and \citet{bat08}
to systematize and optimize the precision of the CaT-based methods.
The CaT is quite sensitive to surface gravity as well as metallicity
(e.g., \citealt{jon84}; \citealt{jor92}),
and so a proxy for this quantity, usually a photometric measure for
practical reasons, is used to obtain a metallicity from the CaT
line width. The main difference between various groups and authors 
is in their choice of passband and reference level for the gravity
proxy.  The most popular choice, dating to the work of \cite{rutl97a},
is the $V$ magnitude distance above the horizontal branch, but the 
$I$ band absolute magnitude has also been advocated \citep{car07}. 
To extend the previous work of \citet{cole04} and measure abundances
and radial velocities for some understudied clusters, we observed
a sample of ten Galactic open clusters and two Galactic globular clusters
with the multifibre spectrograph at the William Herschel Telescope
on La Palma.  In selecting targets, we found it difficult to
obtain $V$ and $I$ photometry across the large spectrographic fields,
even though most clusters have been well-surveyed in their central
regions.  We
realized that the 2MASS catalog was the only homogeneous dataset to 
completely cover all of the clusters in our sample. To preserve as much 
homogeneity as possible, we investigate the CaT - metallicity relationship 
using 2MASS $K_s$ photometry.

We observed a sample of ten Galactic open clusters and
two Galactic globular clusters with the multifibre spectrograph at the
William Herschel Telescope on La Palma.  Five of the ten open clusters (NGC 6791, NGC
6819, NGC 6939, NGC 7142, and NGC 7789) and both of the globular clusters (M15
and M71) have been extensively studied.  Each has had metallicity determinations
from high- or medium-dispersion spectroscopy and model atmosphere analysis, although the 
measurements are diverse in their choices of atmospheres, temperature and 
surface gravity determinations, so the [Fe/H] values are strongly inhomogeneous.  We discuss the
choice of reference metallicities below.  Individual cluster members that have
previously published radial velocities are also used as cross-correlation templates
in our radial velocity measurements.  

In order to reasonably sample the 
line-strength--magnitude--metallicity parameter space, we combined our sample with the 
ten calibration clusters of \citet{cole04}.  2MASS $K_s$ photometry was found for all but three stars
in their sample (NGC 2298: SH156 and SH172 (T. Smecker-Hane, unpublished); NGC 104: L5528 \citep{lee77}). 
Including measurements taken from different
sources can introduce hidden systematic effects if proper care is not taken.  
Nonetheless these seventeen clusters serve as a calibration sample from which we derive our relationship
between CaT equivalent width, $K_s$ magnitude, and metallicity.  We test 
our ability to combine the samples below.  The pertinent cluster parameters are
summarized in Table \ref{clust1}. 

\subsection{Understudied Clusters}

We applied our calibration to five clusters for which we could find no
published spectroscopic metallicity measurement.  Apart from 
northern hemisphere early summer visibility, our only requirement was for the clusters
to contain at least 30 red giants\footnote{as per the listing at the WEBDA
database, http://www.univie.ac.at/webda/}.  Our understudied cluster sample
consists of Berkeley~81, Berkeley~99, IC~1311, King~2, and NGC~7044.
The selected clusters were 
taken to be representative of the surviving old ($\geq$ 10$^9$ yr) open clusters of the 
Milky Way, typified by, e.g., M67, NGC~2141, or the well-studied clusters
of our sample. Some references to IC~1311 make it younger than 10$^9$ yr
(e.g., \citealt{car94} and the WEBDA database value),
but the CMD morphology suggests it is comparable in
age to Be~81 or NGC~7044.
The useful cluster parameters are given in Table \ref{clust2}.  

One of the biggest uncertainties in determining a proper CaT versus metallicity
relationship is the lack of a universally-accepted, high-dispersion fiducial sample of 
globular and open clusters.  Large databases of high-dispersion metallicities have
been created for globular clusters (e.g., \citealt{cg97}; \citealt{ki03}),
and medium-resolution metallicities \citep{friel02} or medium-band photometry
\citep{twa97} for open clusters,
but no large studies combining
globular and open clusters on a homogeneous basis have been made as of yet
(as noted by \citealt{cole04}; \citealt{car07}).
Therefore the metallicities we determine are always referred back to a particular 
metallicity scale from the literature.

\subsection{The Red Giant Star Sample \label{sample-sec}}

Red giant branch (RGB) stars were chosen from 2MASS $J$ and $K_s$ photometry in the regions of the 
selected clusters.  ($K_s,~J-K_s$) CMDs were created for
square-degree areas centered on each cluster, and targets were selected from
the cluster RGB locus down to and including the RGB clump (RC).  The cluster
radii were initially chosen to match the values in \citet{lyn87}, although they
were varied slightly to optimize the contrast between the cluster and the field
in the CMD in some cases.

We tried to sample as wide a range of magnitude
as possible in each cluster in order to accurately model the influence of surface
gravity on the CaT equivalent widths.  The spectrograph fibres are assigned
according to an optimization algorithm that responds to user-supplied weightings
for each star in the target file.  The weightings are required because not 
every desired target can be observed in a single configuration due to the 
requirement that fibres not collide or bend at sharp angles.
We assigned high weights to stars within the adopted cluster
radii and very low weights to those outside; this was necessary because configurations
with a high central concentration of targets are extremely susceptible to fibre
collisions and tend to be avoided by the software.  Because the RGB luminosity
function is steeply
decreasing with brighter magnitude, we weighted bright stars more heavily than stars 
near the RC; we expected that this would prevent the bright stars necessary
to define our surface gravity correction from being excluded, and that an
adequate sample of fainter
stars near the RC would be filled in between the bright star targets with relative
ease.  This procedure appears to have succeeded for all clusters but Be~81, where
a too-broad initial colour selection,
poor tuning of the weighting algorithm and bad luck produced a situation where cluster 
membership remained ambiguous even considering the radial velocity information (section
\ref{be81-sec}).

The central Be~81 problem is a lack of bright RGB stars, but this is an issue
to varying degrees for nearly all open clusters \citep[e.g.,][]{can70}.
Additionally, the brightest, coolest stars are often contaminated
by titanium oxide bands in the spectral region of interest, making them 
unusable for the CaT method.  M stars so identified are noted in the figures
for each cluster.
In most cases, the cluster RGB sequences were not unambiguously 
distinct from the surrounding field, and the samples were cleaned after
observation according
to radial velocity and position relative to the cluster center (see below).  Because
the field of view of the multifibre spectrograph is far larger than the sizes of the 
clusters, a large number of stars that were expected to be field stars were measured
in order to solidly establish values of the foreground/background radial velocity as
an aid to membership distinctions.

\subsubsection{Defining the Red Clump $K_s$ Magnitudes}

It was necessary to define the mean $K_s$ magnitude of the cluster red clumps
(horizontal branches in the case of the globular clusters) so that the height
of each target star above the clump could be defined.  Each of the open clusters
had previous red clump magnitude determinations from \citet{col01}.  The 2MASS
point source catalog ($K_s,~J-K_s$) CMDs for one square degree regions centered
around each cluster were examined, and the radius was gradually reduced until 
further contraction began to exclude red clump stars without reducing the extent
of the clump in $K_s$ or $J-K_s$.  The final selection regions ranged from 
1$\farcm$8 (IC~1311) to 12$^{\prime}$ (NGC~7789); these radii are not identical
with cluster radii in the literature \citep[e.g.,][]{sag98}, but are roughly
similar, and in all cases yielded a well-defined red clump.
The same procedure was followed for the globular clusters, except for M15 and
NGC~4590, for which the horizontal branch was close enough to the limit of the
2MASS photometry that an accurate $K_s$ magnitude could not be found.  In
these two clusters, the appropriate values were taken from \citet{fer00}.
The horizontal branch is not horizontal
in the ($K_s,~J-K_s$) colourspace; for clusters with an extended HB, we 
took the value of $K_s$ at the RR~Lyrae instability strip as the reference
level.
The values of $K_{s,RC}$ for previously observed and understudied clusters
are found in Tables 1 and 2, respectively.

The exact sample selection to define $K_{s,RC}$
does not strongly influence our results in any case.
Independent checks on our derived $K_{s,RC}$ values can be found in 
\citet{gro02} (GS02) and \citet{van07} (vHG07), who each determined
$K_{s,RC}$ for open cluster samples for use as a distance indicator.  
We have five clusters in common with GS02:
NGC~6791, NGC~6819, and (from Cole et al.\ 2004) M67, Be~39,
and 47~Tuc.  The mean difference (us$-$GS02) is $-$0.008~mag, and the largest
difference is $-$0.04 $\pm$ 0.11 mag, for M67.  The same clusters plus NGC~7789 and
Melotte~66 were included in the sample of vHG07; here we find a larger
difference, with (us$-$vHG07) = $-$0.05~mag, ranging from $-$0.03 to $-$0.08.
Shifts this small are entirely insignificant to our derived
metallicities.  Even for Be~81, where just a few red clump stars
within a 3$\farcm$6 radius ($\sim$33\%
larger than the cluster radius from \citet{sag98}) are distributed in
a near-vertical feature,
the adopted error of $\pm$ 0.4~mag on $K_{s,RC}$ only contributes
$\pm$ 0.06 to the error budget for [Fe/H] (see below).

\subsection{Data Acquisition and Reduction}

The data were obtained on 27--28 June, 2004 at the 4.2m William Herschel
Telescope at La~Palma, Spain.  We used the WYFFOS wide field fibre optical
spectrograph, fed by the AF2 fibre-positioning system that allows simultaneous
measurement of up to 150 objects over an un-vignetted 40$^{\prime}$ diameter
field of view through its 1$\farcs$6 diameter fibres.  We used the low-order
`echelle' grating in 3rd order with $\lambda _{cen}$ = 8701 \AA\ and the GG495
order-blocking filter, yielding
a spectral coverage of 570 \AA\ and a spectral resolution of $\approx$ 0.15
\AA/pix.  The weather was clear but the seeing was moderate to poor, resulting
in some light loss from the fibres.  Typical exposure times ranged from 480--900
seconds.  Offset sky exposures were taken after each science frame, with a typical
offset of 2 arcseconds.  Neon arclamp and screen flat exposures were also taken
subsequent to each image to allow flatfielding and wavelength calibration.

Data reduction was performed under IRAF using Pierre Leisy's instrument-specific
reduction scripts\footnote{http://www.ing.iac.es/Astronomy/instruments/af2}.  The
reduction package corrects the CCD bias using zero frames and an overscan region,
and trims the image.  The individual fibres are automatically found, traced along
the CCD, and extracted using standard flatfield and wavelength calibration procedures.

Examination of the residuals from the sky subtraction process showed that the 
night-sky emission lines of OH \citep{ost92} were frequently either under- or 
over-subtracted during the calibration.  Therefore we developed an iterative process
to optimize the sky subtraction for each fibre in each image.
We processed the offset sky fields for each cluster using the same reduction
script parameters as used for the clusters themselves, and matched each fibre
in the offset sky to the corresponding fibre in the science exposure.  Each
sky spectrum was scaled by a wavelength-independent factor of order 
unity and subtracted from the matching object spectrum, leaving a residual spectrum
whose rms scatter was measured.  The scaling factor was then adjusted iteratively
until the rms reached a minimum value.

\begin{figure}
\includegraphics[width=168mm]{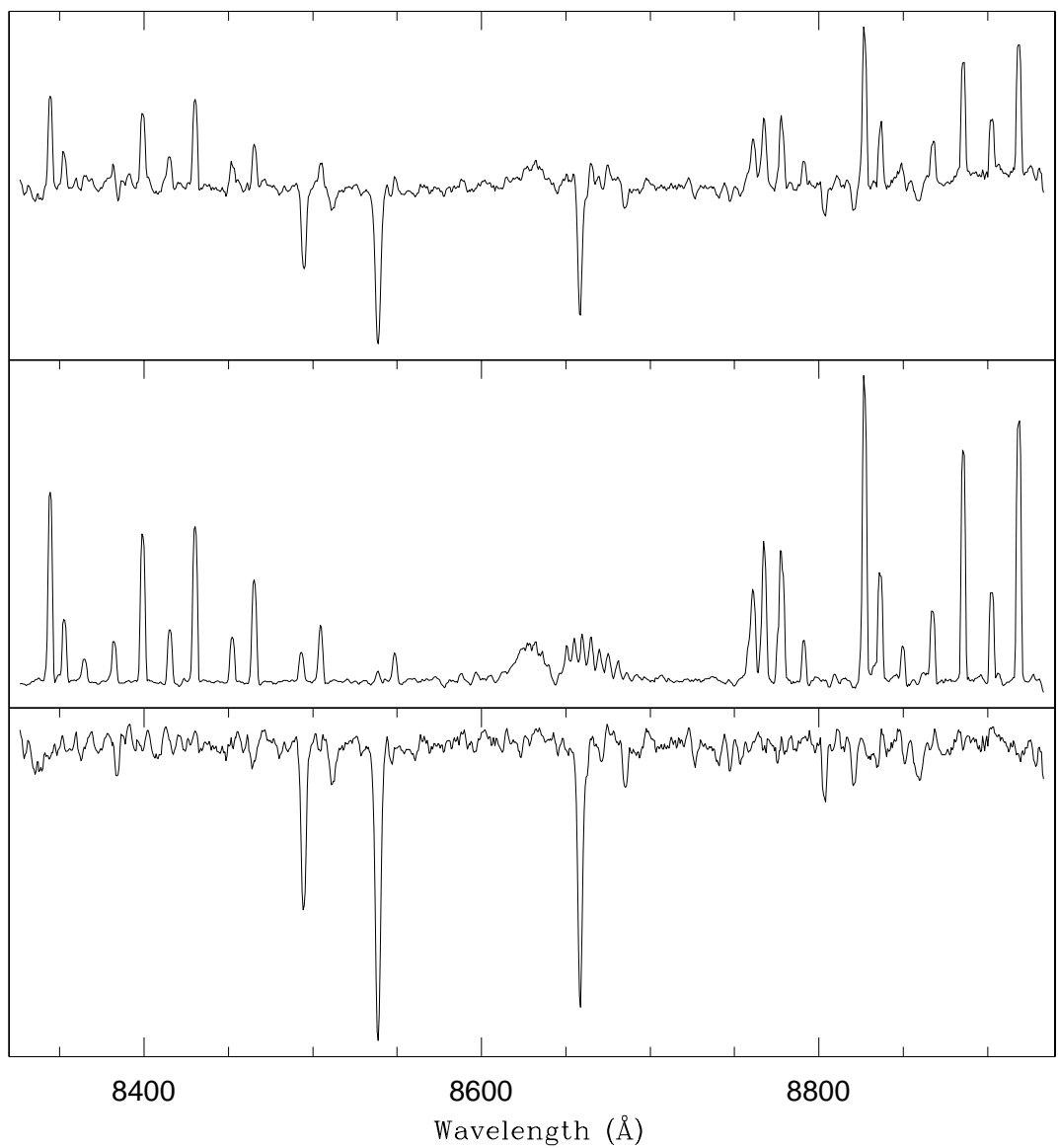}
\caption{The raw spectrum, offset sky, and reduced spectrum (continuum normalized) for a typical target star
(M15 \#21300038+1207363). \label{fig-sample}}
\end{figure}

The sky-subtracted exposures were then continuum normalized using a low-order polynomial
fit to spectral regions excluding areas of strong telluric water vapor absorption.
Figure \ref{fig-sample} shows the raw, sky, and reduced spectra for a typical target.
Except in a few cases, the S/N/pixel in our final, extracted spectra was $\gtrsim$ 20.

\subsection{Radial Velocities}

\begin{figure}
\includegraphics[width=168mm]{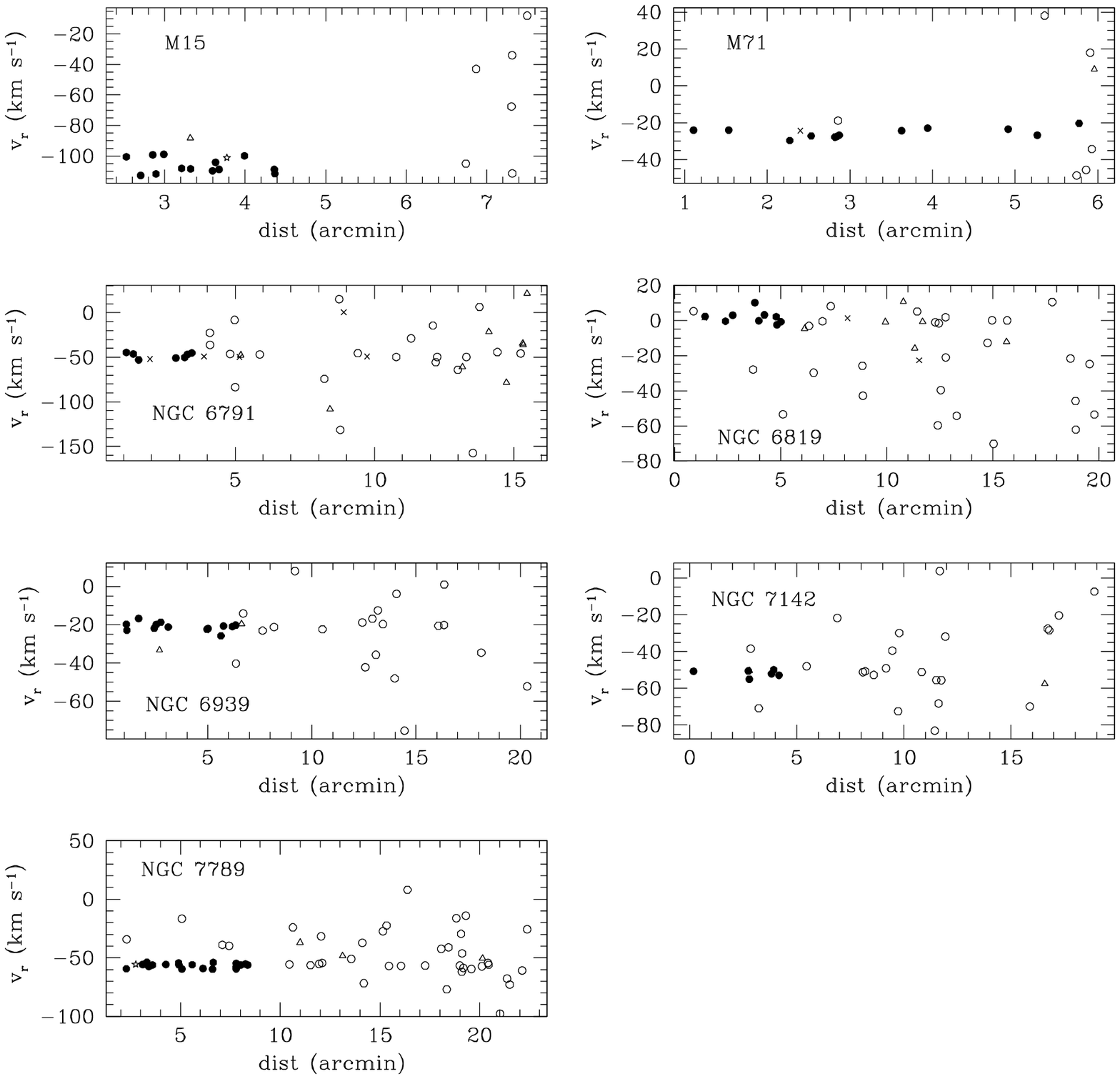}
\caption{Filled black circles are the accepted cluster stars based upon radial velocity
and positional determinations.  Open symbols are stars rejected from further analysis based
upon radial velocities/falling beyond the visual cluster bounds (circles), TiO bands
(crosses), CN bands (stars), and/or low quality spectra (triangles).
\label{rvdist}}
\end{figure}

Radial velocities are important in determining membership probabilities as well
as estimating the expected line centers of the CaT for use in the equivalent
width measurements.  We derived relative radial velocities of our sample stars
by cross-correlating their spectra with reference stars within our sample
\citep{ton79}.  The
reference stars chosen have previously derived radial velocities taken from the
literature.  We chose three stars from M15 \citep{sod99}, three from NGC 7789
\citep{gim98}, and one from NGC 6819 \citep{friel93}. The IRAF FXCOR task was 
then used to perform the cross-correlation
with each of the seven reference spectra.  Final radial velocities were then
derived by taking the weighted average of the results from each reference
spectrum.  The weighted average included the errors in the correlation
and correlation peak height.  Stars whose radial velocities differed greatly
from the cluster mean radial velocity were removed from consideration as well as
those stars whose distances from the cluster center were beyond visually
inspected cluster bounds.  The velocities and standard errors are given 
in Table \ref{tab-stars} and the velocity cuts are shown in Figure \ref{rvdist}.

\subsection{Equivalent Width Measurements}

Methods for deriving the equivalent widths of the CaT lines have been previously
discussed in \citet{cole04}.  \citeauthor{cole04} showed that a sum of a 
Gaussian and a Lorentzian produced an empirical function that fit the line
profiles more accurately over a wide range of line strengths than a Gaussian
(e.g. \citealt{arm91}) alone.  The
Gaussian profile, while having an acceptable fit for weak lines, underestimates
the equivalent widths for strong lines by failing to incorporate the broadened
CaT wings.  The same is true for the Moffat function used by \citeauthor{rutl97a}
as shown by \citet{pont04}.  Based on this, we have used the sum of a Gaussian 
and Lorentzian to define our empirical function and integrated the result to 
calculate the equivalent width of the CaT lines.  We measured the equivalent 
widths using the {\tt ew} program \citep{cole04} to fit the continuum around 
each line with a linear function and the lines with a Gaussian$+$Lorentzian
that are forced to have the same central wavelength.  Bad pixels and residuals from
cosmic ray removal are rejected from the line and continuum fits using an iterative
sigma-clipping method, and the residuals to the fits are visually inspected.
The line and continuum bandpasses are listed in Table \ref{bpass}.

The equivalent width of a spectral line can be strongly affected by other atomic
and molecular line opacities that alter the continuum level around the features
of interest.  Two important contaminants around the calcium triplet are titanium
oxide (TiO) and cyanogen (CN) bands.  TiO can suppress the continuum between 
8430 \AA\ $\lesssim \lambda \lesssim$ 8550 \AA, making the CaT lines
appear weaker than they actually are.  TiO bands become problematic for giants
cooler than late K-type, and are especially troublesome at high metallicity 
(e.g., NGC~6791).  CN bands can have a similar depressing effect on the CaT lines,
but they are more symptomatic of stellar nucleosynthesis and dredge-up processes
than of cool temperatures.  We have carefully inspected the spectrum of each star
and rejected from further analysis the targets that showed the signatures of TiO
or CN bands.

\subsubsection{The CaT Index}

The CaT index ($\Sigma$W; \citealt{arm91}) is taken as a linear combination 
of the pseudo-equivalent widths of the three Ca~II lines ($\lambda\lambda$ = 
8498.02, 8542.09, 8662.14 \AA).  \citet{rutl97a} explored the impact of 
including or excluding the weakest line and various weighting schemes as
a function of signal-to-noise.  Studies with poor spectral resolution and/or
low S/N usually form $\Sigma$W excluding the weak 8498\AA\ line \citep{sun93,cole00,bat08},
while others have used an unweighted sum of all three lines \citep{ols91,cole04,car07}.
Given the high quality of our spectra (S/N/pixel $\gtrsim$ 20) we have adopted
a straight sum of the three individual lines:

\[
\Sigma W  = W_{8498}  +  W_{8542}  +  W_{8662}.
\]
The exact value of $\Sigma$W will vary from study to study because there is some
freedom in choosing the continuum level to which the equivalent width values
are referred.  The values should fall within the uncertainties reported for 
$\Sigma$W if the errorbars are properly estimated.

\citet{arm91} introduced a `reduced' equivalent width parameter, W$^{\prime}$, used to
effectively remove the surface gravity and effective temperature dependences of the 
line strength, leaving only the metallicity dependence.  Because red giants
lie along a narrow sequence in the luminosity (surface gravity) vs.\ temperature plane,
T$_{eff}$ and log($g$) are correlated with each other and their influence on $\Sigma$W
can be calibrated out using a single observable.  Colour and absolute magnitude have
both been used in the past to create the index W$^{\prime}$, but the most robust method
in the presence of distance and reddening uncertainties is to use an expression relating
the magnitude of the target star to the mean magnitude of the horizontal branch (or 
red clump) of its parent population:
\[
W^\prime = \Sigma W + \beta (\mathfrak{M}-\mathfrak{M}_{HB}),
\]
where $\beta$ depends on the definition of $\Sigma$W and on
the bandpass $\mathfrak{M}$ used in the comparison.  Since we assume that each star 
in a given cluster has the same metallicity and thus the same W$^{\prime}$ value, 
the error on W$^{\prime}$, $\sigma_{W^{\prime}}$, depends on the standard deviation of 
the individual stars' W$^{\prime}$ values about the mean cluster value as well as the 
number of stars, N, used in each cluster.
\[
\sigma_{W^{\prime}}~=~\frac{\sigma}{\sqrt{N}}
\]

Because much of the opacity
in the very strong Ca~II lines lies in the wings, the measured increase of line strength
with decreasing surface gravity is to some degree dependent on the functional form
adopted to measure the lines, and on the S/N of the data.  Therefore the standard
best practice is for calibration clusters to be observed, providing a constant
metallicity sample from which appropriate $\beta$ values can be derived. Individual studies
have found no dependence of $\beta$ on metallicity or age \citep{rutl97a},
but studies including open as well as globular clusters tend to find slightly
higher $\beta$ values (e.g., Cole et al.\ 2004).  If stars below the horizontal
branch or above the tip of the red giant branch are included, the linear relation
between $\Sigma$W, $(\mathfrak{M}-\mathfrak{M}_{HB})$, and [Fe/H] breaks down,
and an $(\mathfrak{M}-\mathfrak{M}_{HB})^2$ term must be introduced \citep{car07}.

Previously, Johnson $V$ and Cousins $I$ have been the most commonly used magnitudes
for the formation of W$^{\prime}$.  Typical values are
$\beta_V$ $\approx$ 0.7 (e.g., Rutledge et al.\ 1997b, Cole et al.\ 2004).
However, in many open clusters only a small
region of the cluster center has published photometry, and in some cases that
photometry predates the CCD era and so would be difficult to replicate consistently.
However, the Two Micron All-Sky Survey has provided near-infrared $JHK_s$
magnitudes on a uniform scale over the entire sky down to $K_s$ $\approx$ 14, below
the red clump/horizontal branch level for many open clusters.  This provides
the tantalizing prospect of using the near-infrared photometry to form W$^{\prime}$,
obviating the need to search through a large number of inhomogeneous studies
to accumulate the necessary $VI$ photometry.  

Using 2MASS data instead of optical data 
also has many advantages.  $K_s$ is less sensitive to reddening by dust than $V$ or $I$, 
therefore any intra-cluster reddening dependence in W$^{\prime}$ is reduced.
Also, \citet{carp01} derived transformation equations for many infrared filter sets
such that their $JHK_s$ colours can be placed on the 2MASS system.  This gives
a unique opportunity to maintain homogeneity in future studies.

\subsection{W$^{\prime}$ and Metallicity}

Typically, a linear relationship has been established between W$^{\prime}$ and [Fe/H],
but this depends crucially on the adopted metallicities of the calibration sample (see
\citealt{cole04} and references within).  A linear relation between W$^{\prime}$ and
[Fe/H] has been called into question for metallicities on the widely-used \citet{zw84}
scale as well as high-dispersion scales based on measurements of ionised rather than
neutral iron \citep{ki03}.  Adopting metallicities from
different sources will change the derived slope and intercept values.  Higher
order terms in the W$^{\prime}$ vs.\ [Fe/H] relationship
(\citealt{car01};\citealt{cole04};\citealt{car07}) may then be artificially introduced 
(see \S\ref{diffscale}).  With these issues in mind, we have decided to use the metallicities defined 
by \citet{cg97} for the globular clusters and \citet{friel02} for the open clusters.
Our best hope is to try and remain consistent throughout our analysis.  

\subsection{Combining the Two Data Sets}

Before we can combine our cluster sample with that of \citet{cole04} we must first 
test if they are compatible with eachother.  One way to do this is to derive the W$^{\prime}$
vs. [Fe/H] relationship for each sample individually and compare the results.  Systematic
effects from instrumental and reduction differences can introduce non-linearities if the 
samples are incompatible.

In order to derive $\beta$ for an individual cluster, we used 2MASS $K_s-K_{s,RC}$ 
magnitudes for each star
plotted against $\Sigma$W.  The slope of the best fit line, weighted by errors
in $\Sigma$W, was computed for each calibration cluster.  The mean slope value,
$\beta_{K_s}$, weighted by the errors in the slopes,
was then calculated.  By using the 2MASS 
$K_s$ magnitudes we derive a value of $\beta_{K_s} = 0.45 \pm 0.03$ for our sample and 
$\beta_{K_s} = 0.49 \pm 0.02$ for the \citet{cole04} sample.  

\begin{figure}
\includegraphics[width=168mm]{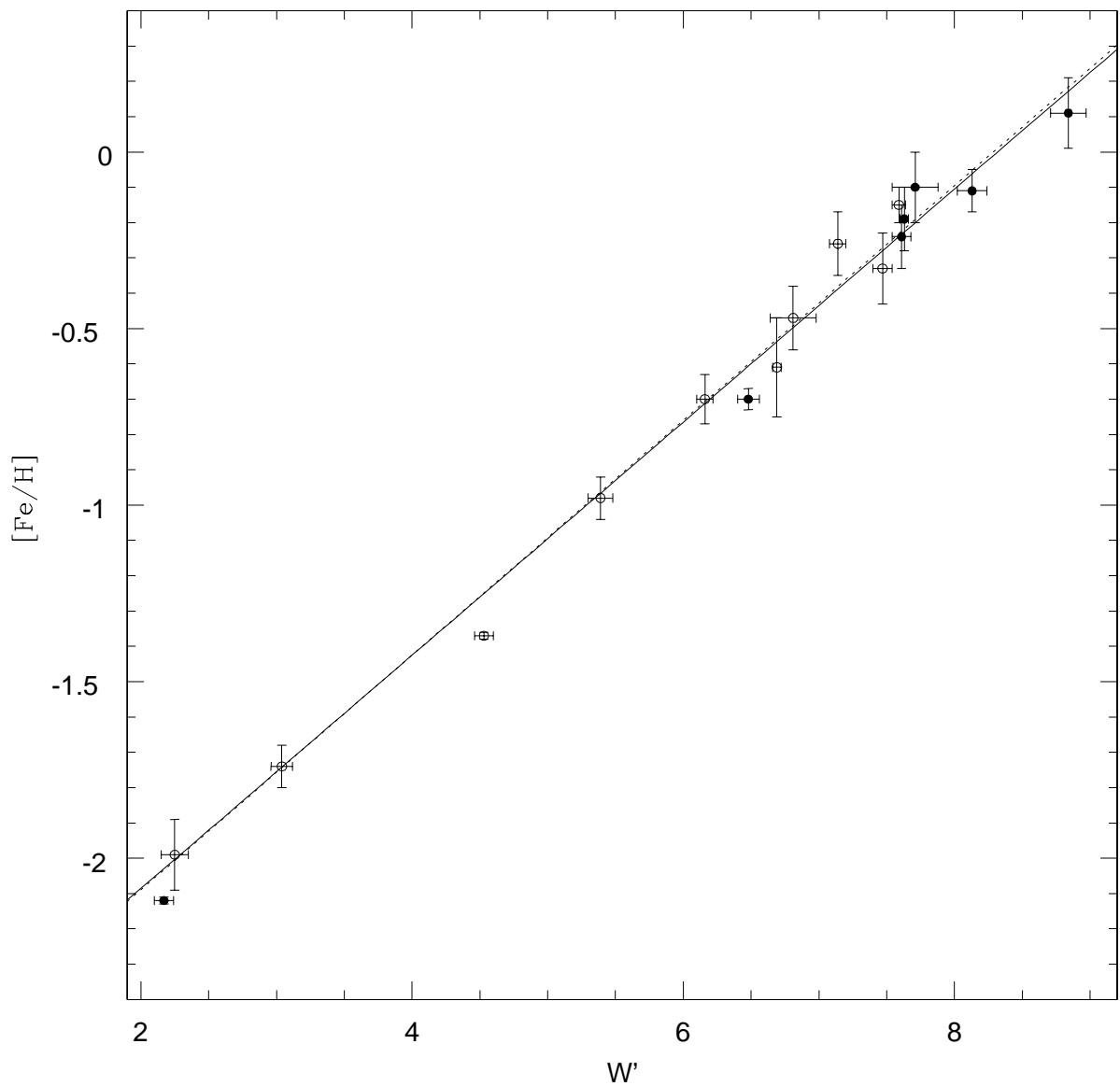}
\caption{Filled circles and the best fit solid line are for our cluster sample.  Open circles and the best fit dashed 
line represent the clusters of the \citet{cole04} sample.\label{wpcomp}}
\end{figure}

We then derived the average W$^{\prime}$ value for each cluster and plotted them against the
reference metallicities given in Table \ref{clust1}.  We then fit each cluster sample with a line 
weighted by the metallicity errors (see Figure \ref{wpcomp}).  The equations of the best line fits are 
\[
[Fe/H] = (-2.746~\pm~0.224)+(0.330~\pm~0.029)W^{\prime}
\]
for our sample and 
\[
[Fe/H] = (-2.754~\pm~0.076)+(0.332~\pm~0.012)W^{\prime}
\]
for the \citet{cole04} sample.  The two lines are statistically similar and show that the
two samples are indeed compatible.  Any systematic effects between the samples are 
small and cannot be characterised here.

\begin{figure}
\includegraphics[width=168mm]{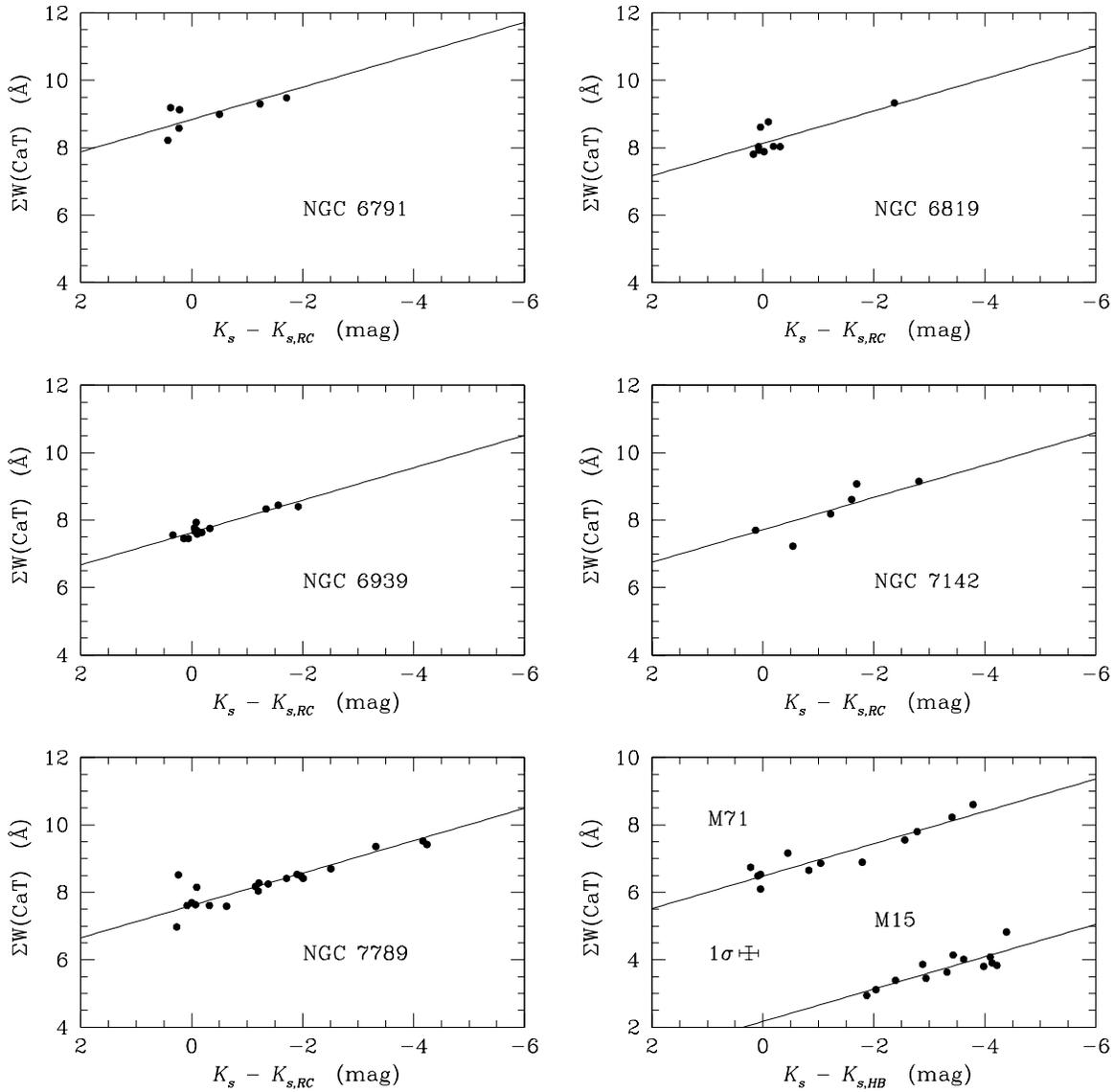}
\caption{The best fit line using $\beta_{K_s}$ = 0.48 as the slope for each of our
calibration clusters.  The typical $1\sigma$ errors for the points are in the bottom
right panel.\label{sewmag}}
\end{figure}

\begin{figure}
\includegraphics[width=168mm]{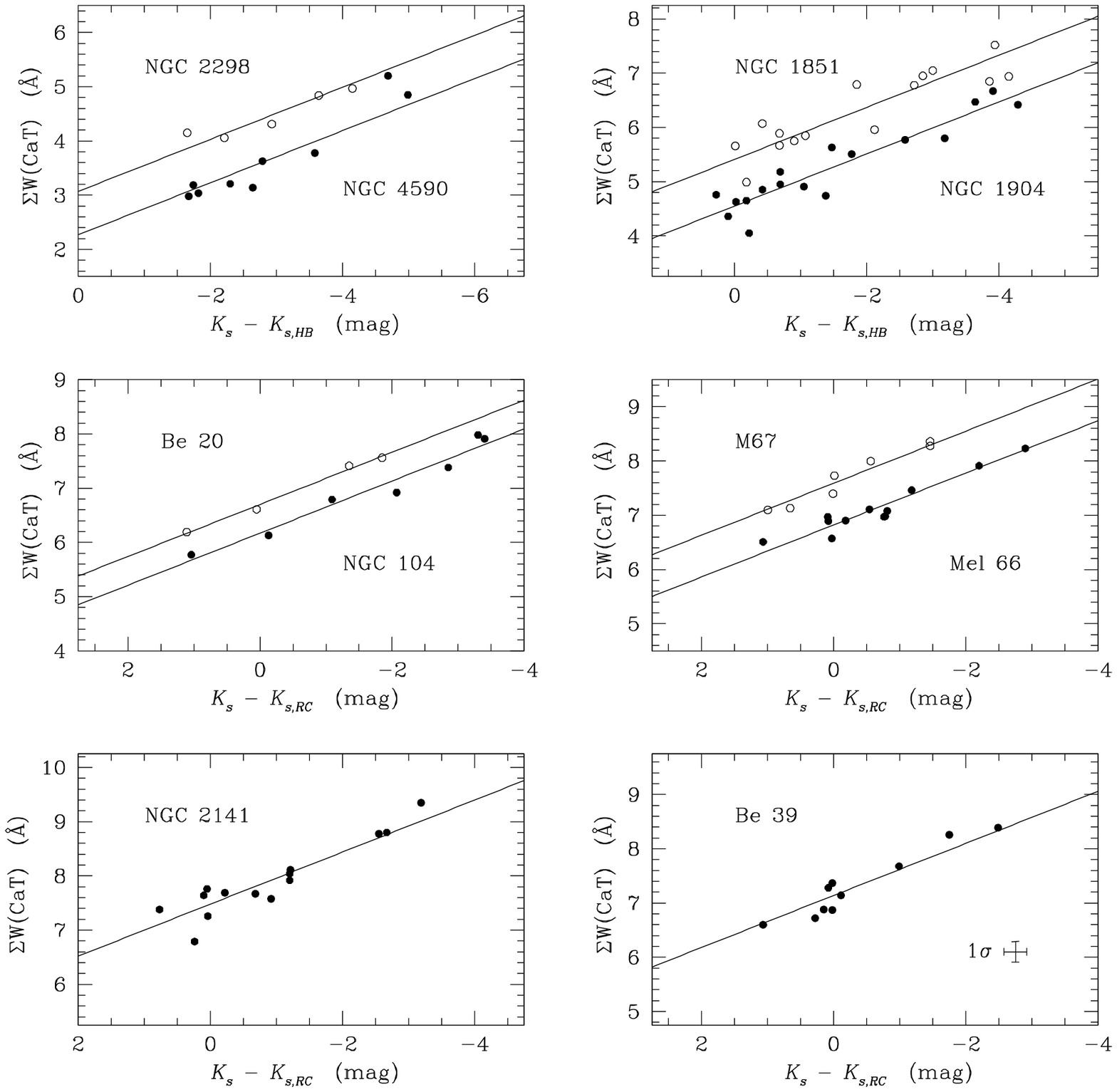}
\caption{The best fit line using $\beta_{K_s}$ = 0.48 as the slope for each of the
\citet{cole04} calibration clusters.  The typical $1\sigma$ errors for the points are in the bottom
right panel.\label{sewcomb2}}
\end{figure}

We can now move forward and combine the two samples and rederive $\beta_{K_s}$ for the 
entire data set.  Using all seventeen calibration clusters we derive a value of 
$\beta_{K_s}$ = 0.48 $\pm$ 0.01.  Figures \ref{sewmag} and \ref{sewcomb2} show the best fit line with this 
slope through each of the calibration clusters.

\begin{figure}
\includegraphics[width=168mm]{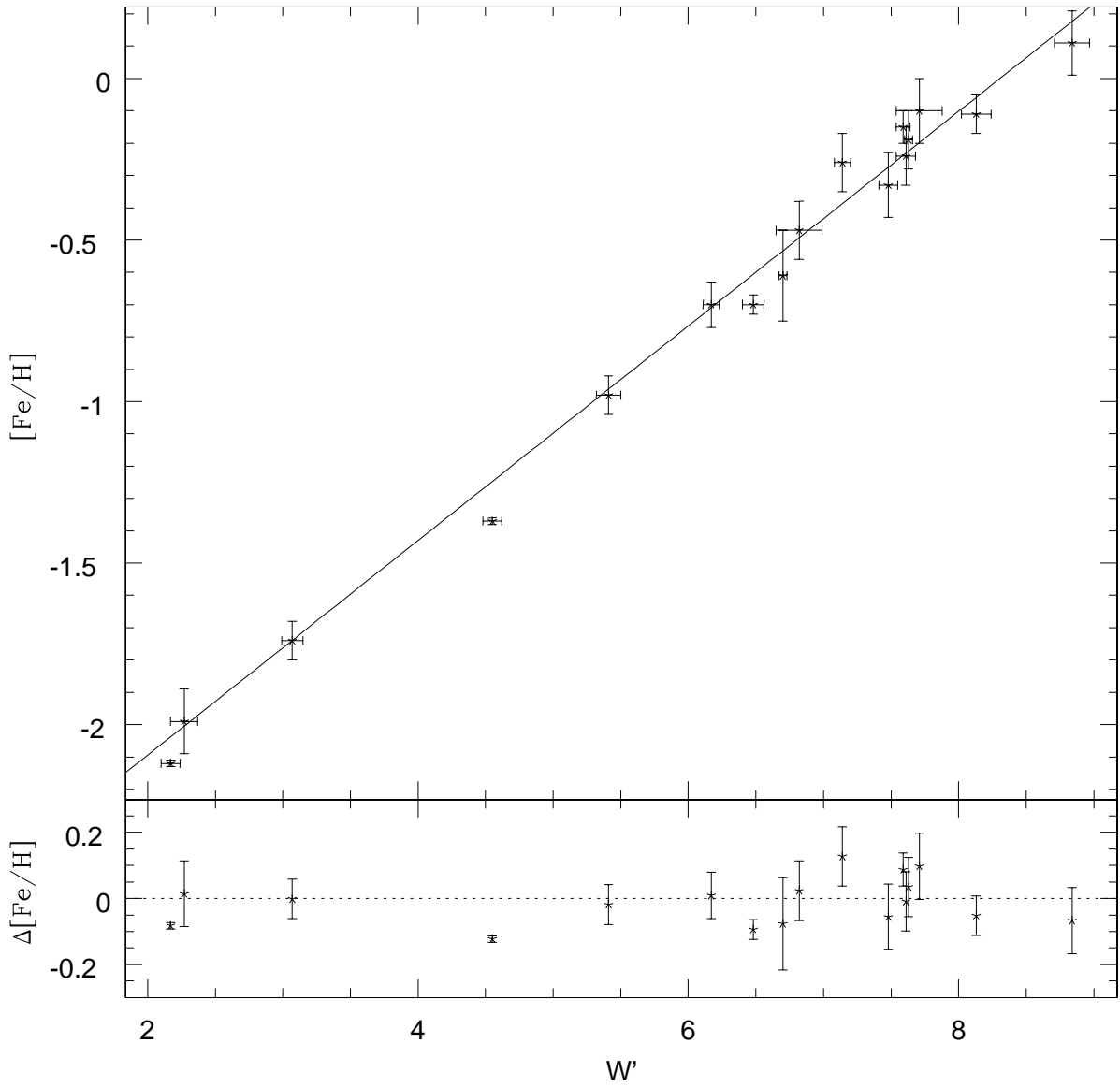}
\caption{Top Panel: Calibration cluster metallicities from Table \ref{clust1}
plotted versus their W$^{\prime}$ values.  The weighted (by metallicity) best fit line 
is also shown.
Bottom panel: Residuals to the best fit line with the 1$\sigma$
reference metallicity errors shown.\label{mvw}}
\end{figure}

With this $\beta_{K_s}$ value, we then derived the final average W$^{\prime}$ value for each cluster and plotted them 
against the reference metallicity values given in Table \ref{clust1}.  
Figure \ref{mvw} shows our best fit line to the data points with our adopted metallicities and
derived W$^{\prime}$ values.  The equation of the fit is
\[
[Fe/H] = (-2.738~\pm~0.063)+(0.330~\pm~0.009)W^{\prime}
\]
which gives consistent predictions with what was derived by \citet{cole04} (see Table \ref{compsC04}). 
Adding a second order term (W$^{\prime 2}$) does not significantly improve the fit.

In order to evaluate our ability to estimate a cluster's metallicity, 
a check of our predicted [Fe/H] value versus the true value is needed. The lower
panel of Figure \ref{mvw} shows $\Delta$[Fe/H] vs. W$^{\prime}$ for each of the
calibration clusters.  The residual about the mean is -0.011 with a standard 
deviation, $\sigma_{scat}$, of 0.07.  The scatter is expected given the scatter in W$^{\prime}$
and the reported errors on reference metallicity, and there are no obvious trends
with cluster age or mean $K_s-K_{s,RC}$.  Final errors on the [Fe/H] predictions depend on the slope
of the derived [Fe/H] line (m$_{[Fe/H]}$), the error on W$^{\prime}$ ($\sigma_{W^{\prime}}$), and $\sigma_{scat}$.
\[
\sigma_{[Fe/H]}~=~\sqrt{(m_{[Fe/H]} * \sigma_{W^{\prime}})^2~+~\sigma_{scat}^2}
\]
Tables \ref{comps} and \ref{compsC04} give our W$^{\prime}$, predicted [Fe/H], and reference [Fe/H]
values.  Table \ref{comps} also gives our derived 
average cluster radial velocities and literature radial velocities values for our cluster sample.  
The derived cluster radial velocities are statistically similar to the literature values.

\subsection{Different Metallicity Scales \label{diffscale}}

\begin{figure}
\includegraphics[width=168mm]{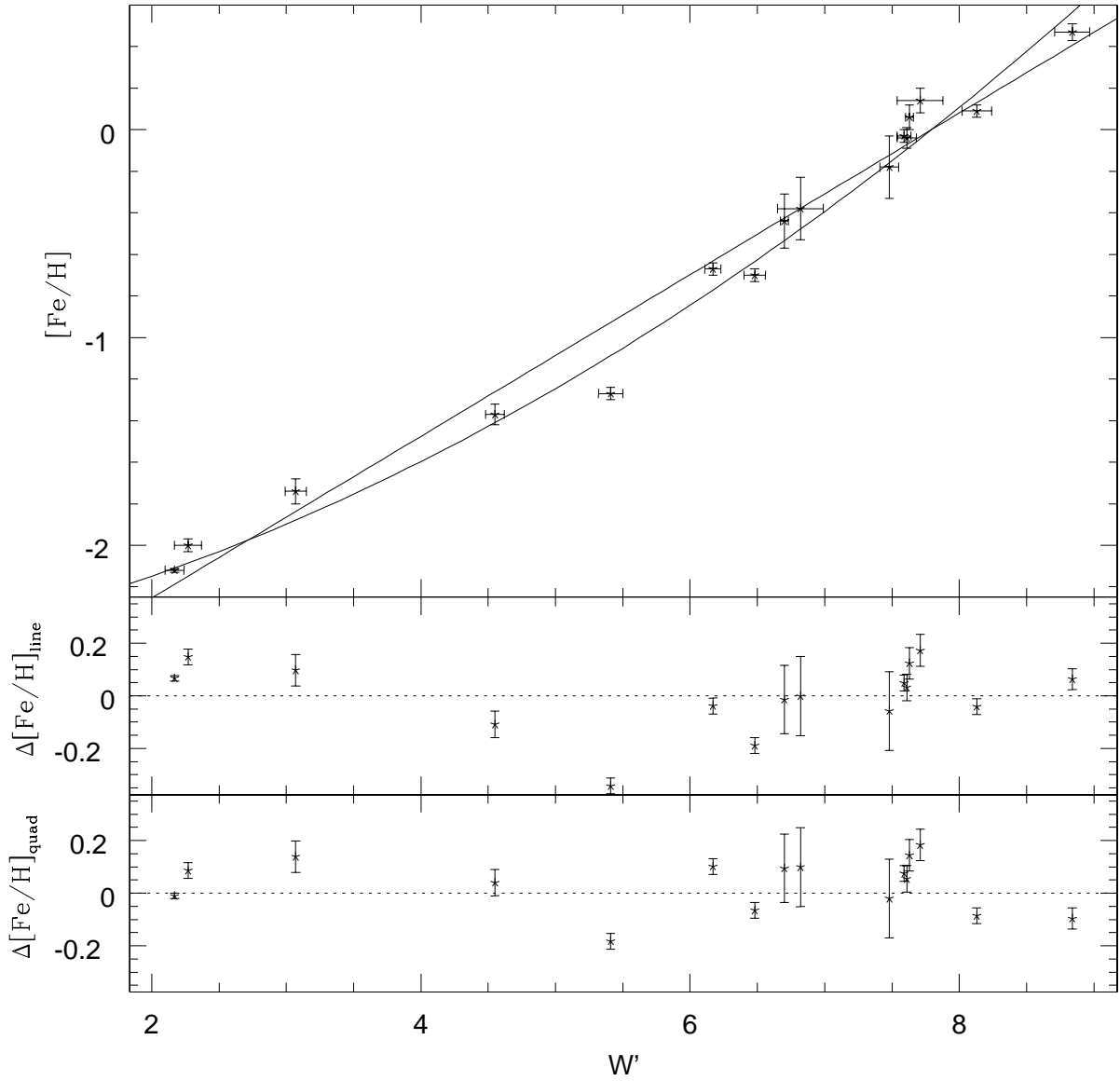}
\caption{Top Panel: Calibration cluster metallicities from Table \ref{dmclust}
plotted versus their W$^{\prime}$ values listed in Tables \ref{comps} and \ref{compsC04}.  
The weighted (by metallicity) best fit line and parabola are also shown.
Middle panel:Residuals to the best fit line with the 1$\sigma$
reference metallicity errors shown.
Bottom panel: Residuals to the best fit parabola with the 1$\sigma$
reference metallicity errors shown.\label{dmresid}}
\end{figure}

If we choose reference metallicites from multiple sources we lose homogeneity in the 
most crucial part of the calibration procedure.  As noted before, no studies have 
produced high-dispersion metallicities for all of our calibration clusters, so the 
final metallicity values are always referred back to the adopted scales.  As an example
to the sensitivity of the metallicity scale on the derived [Fe/H] vs. W$^{\prime}$ 
calibration, we chose metallicity values derived from high-dispersion spectroscopy 
measurements made by different authors.  We took our derived W$^{\prime}$ values for each cluster and 
plotted them against the reference metallicities listed in Table \ref{dmclust}.  
Berkeley 39 currently has no high-dispersion metallicity estimates, so it was left out of this
calibration.  Table \ref{dmclust} and Figure \ref{dmresid} 
give the results of the fits. 

The linear fit in the top panel of Figure \ref{dmresid} has an equation of
\[
[Fe/H] = (-3.031~\pm~0.118)+(0.389~\pm~0.017)W^{\prime}
\]
with the residuals to the fit shown in the middle panel ($\sigma_{scat}$ = 0.13).  The 
quadratic fit results in
\[
[Fe/H] = (-2.501~\pm~0.081)+(0.126~\pm~0.045)W^{\prime}
+(0.025~\pm~0.004)W^{\prime2}.  
\]
The residuals to the quadratic fit are shown in the bottom panel ($\sigma_{scat}$ = 0.10).
Neither fit perfectly reproduces the reference metallicities over the entire range if 
our W$^{\prime}$ errors and the quoted errorbars on the cluster metallicities are realistic
measures of the uncertainty.  Neither
fit appears significantly better than the other, and the significance of any putative 
improvement would be entirely dependent on the exact metallicity chosen for each calibration cluster
and the adopted metallicity uncertainties; i.e., the variation in metallicity between
disparate authors is usually larger than the quoted errorbars. 
This underscores the need for a uniform 
high-dispersion metallicity scale
spanning clusters with a wide range of metallicities.  For the remainder of this paper
we will report the metallicities of our understudied clusters on the linear scale based on
the combined Carretta-Gratton and Friel et al.\ samples, but future workers may recalibrate
our reported W$^{\prime}$ values to any desired scale as new information becomes available.

\section{Results:  New Cluster Velocities and Metallicities}

The results for our calibration clusters show that the reduced equivalent width
of the CaT obtained using 2MASS $K_s$ magnitudes can reproduce the [Fe/H] measurements
obtained from high- and medium-dispersion spectroscopy and CaT measurements plus visible-light
bandpasses.  Therefore we can proceed with some confidence to derive metallicity 
estimates for previously understudied clusters.  Prior to this study, very little was 
known about their metallicities and their radial velocities appeared to be unknown. 
Each of these clusters has at 
least one broadband CCD-based CMD in the literature, but medium-band photometry (e.g., 
DDO or Str\"{o}mgren system) could not be found.

\begin{figure}
\includegraphics[width=168mm]{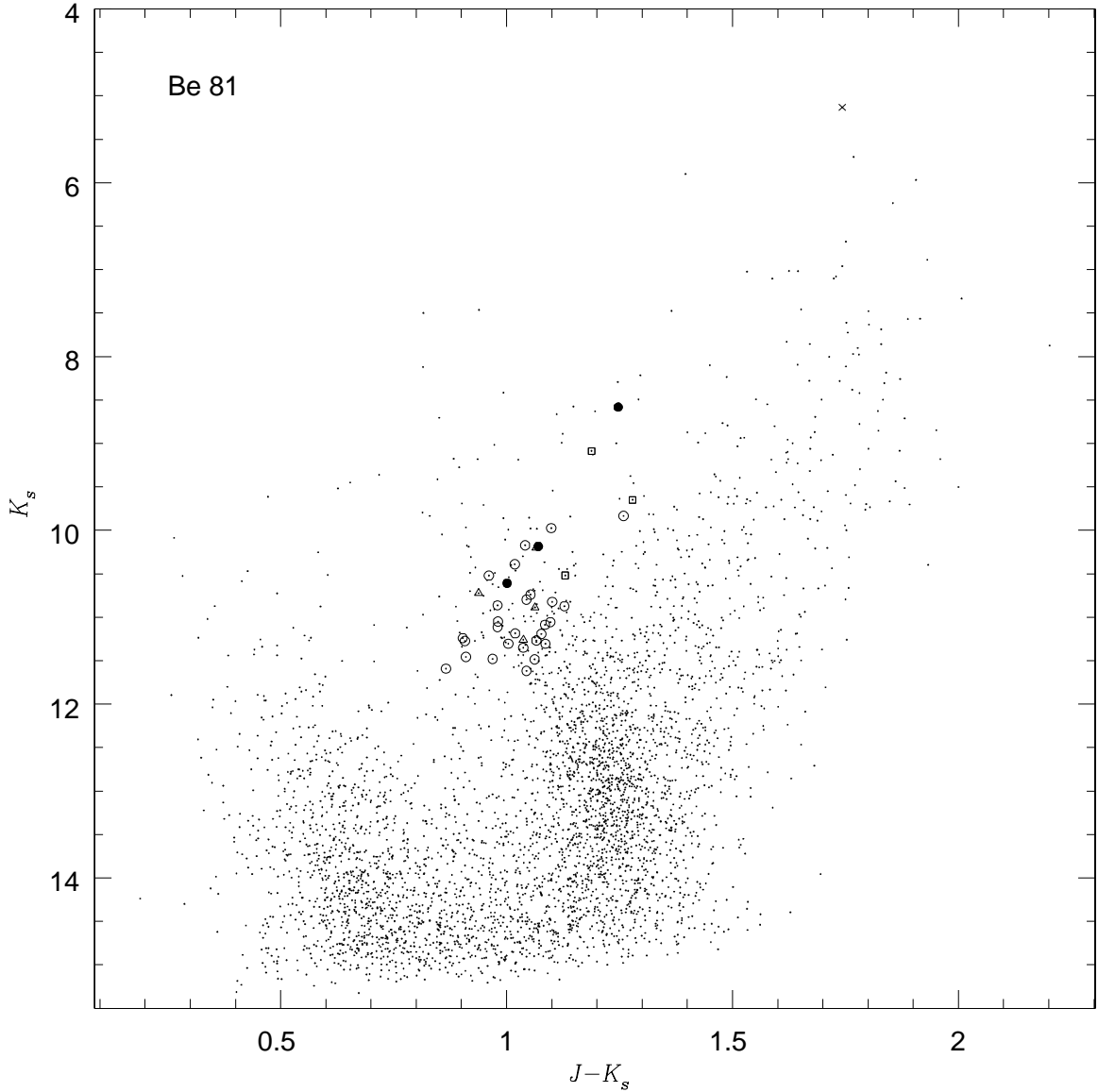}
\caption{The 2MASS ($K_s, J - K_s$) CMD for Berkeley 81.  Filled black circles are the accepted cluster 
stars based upon radial velocity
and positional determinations.  The open squares are part of the ``fainter" subgroup and optical
photometric outlier discussed in \S\ref{be81-sec}.  Other open symbols are stars rejected from further analysis based
upon radial velocities/falling beyond the visual cluster bounds (circles), TiO bands
(crosses), CN bands (stars), and/or low quality spectra (triangles). \label{be81cmd}}
\end{figure}

\begin{figure}
\includegraphics[width=168mm]{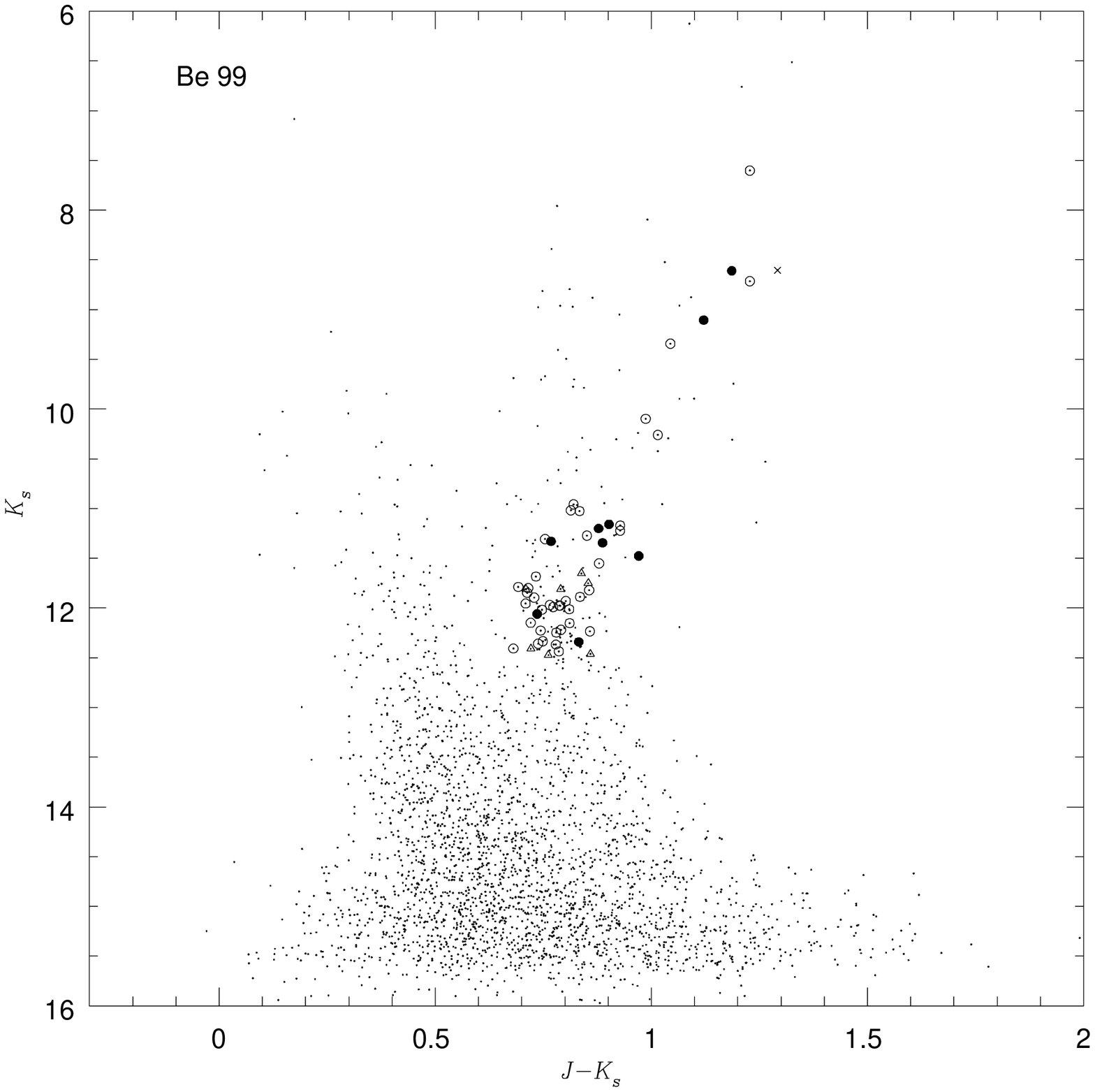}
\caption{The 2MASS ($K_s, J - K_s$) CMD for Berkeley 99.  Filled black circles are the accepted cluster stars based upon radial velocity
and positional determinations.  Open symbols are stars rejected from further analysis based
upon radial velocities/falling beyond the visual cluster bounds (circles), TiO bands
(crosses), CN bands (stars), and/or low quality spectra (triangles). \label{be99cmd}}
\end{figure}

\begin{figure}
\includegraphics[width=168mm]{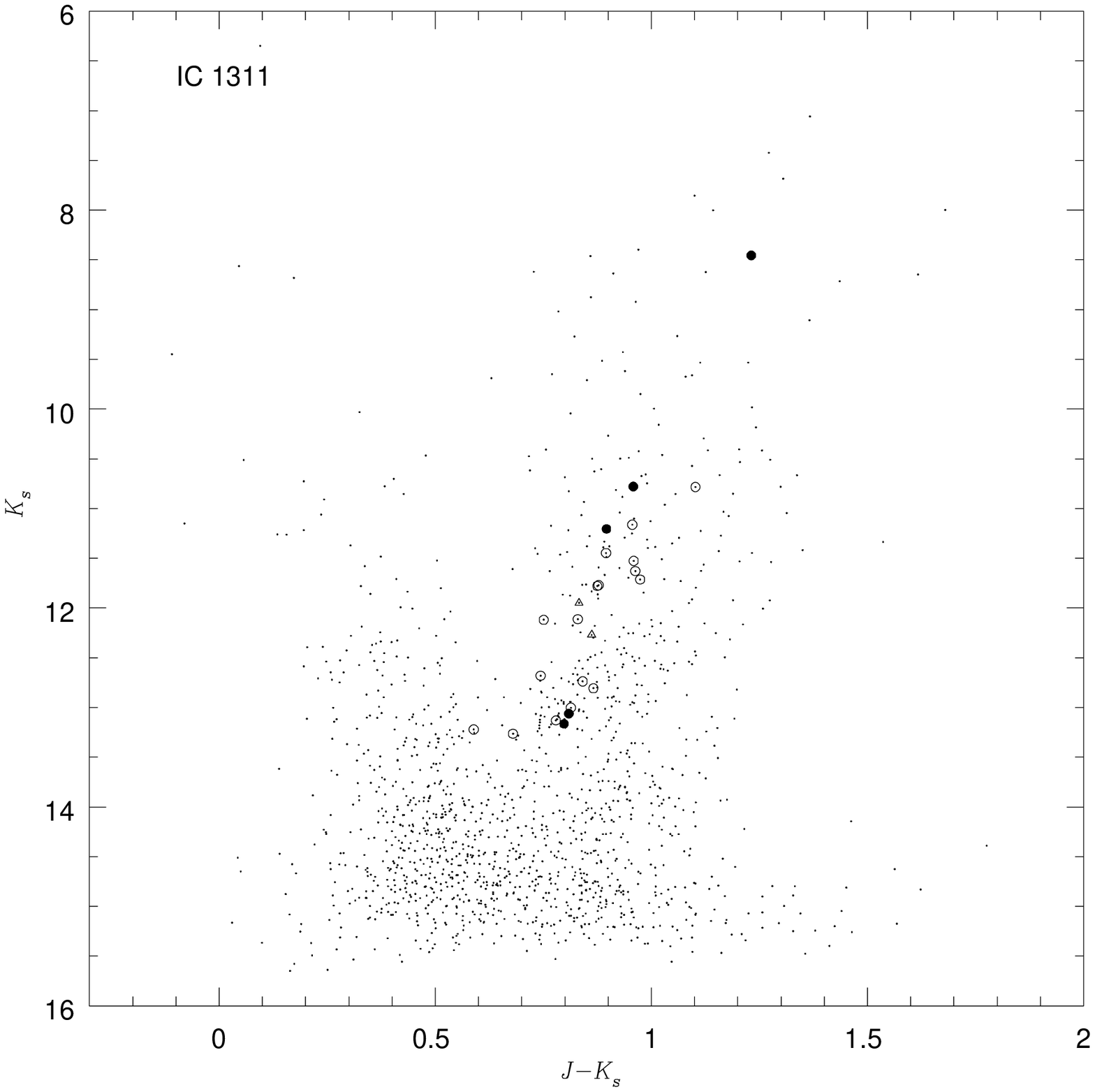}
\caption{The 2MASS ($K_s, J - K_s$) CMD for IC 1311.  Filled black circles are the accepted cluster stars based upon radial velocity
and positional determinations.  Open symbols are stars rejected from further analysis based
upon radial velocities/falling beyond the visual cluster bounds (circles), TiO bands
(crosses), CN bands (stars), and/or low quality spectra (triangles). \label{ic1311cmd}}
\end{figure}

\begin{figure}
\includegraphics[width=168mm]{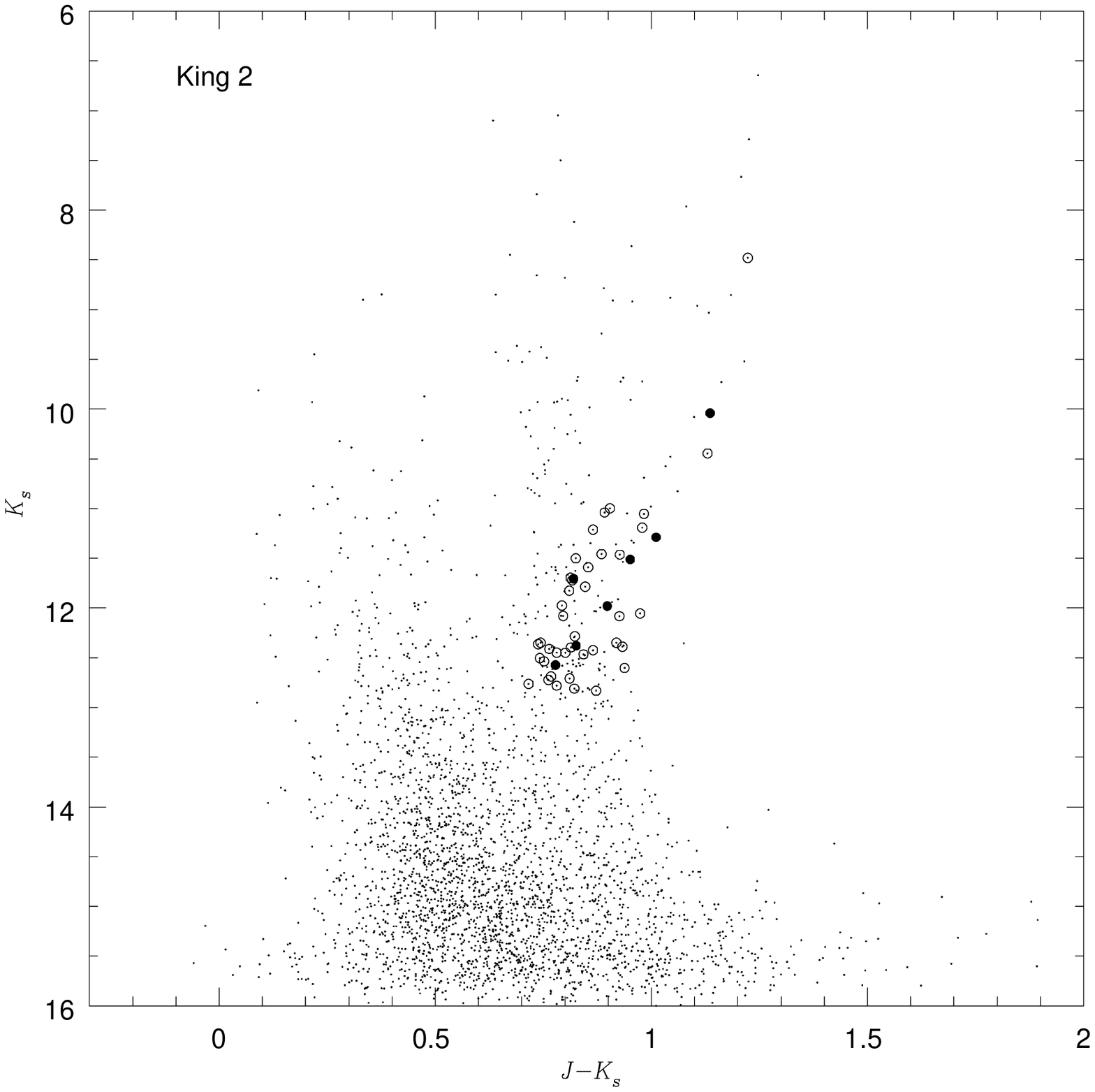}
\caption{The 2MASS ($K_s, J - K_s$) CMD for King 2.  Filled black circles are the accepted cluster stars based upon radial velocity
and positional determinations.  Open symbols are stars rejected from further analysis based
upon radial velocities/falling beyond the visual cluster bounds (circles), TiO bands
(crosses), CN bands (stars), and/or low quality spectra (triangles). \label{king2cmd}}
\end{figure}

\begin{figure}
\includegraphics[width=168mm]{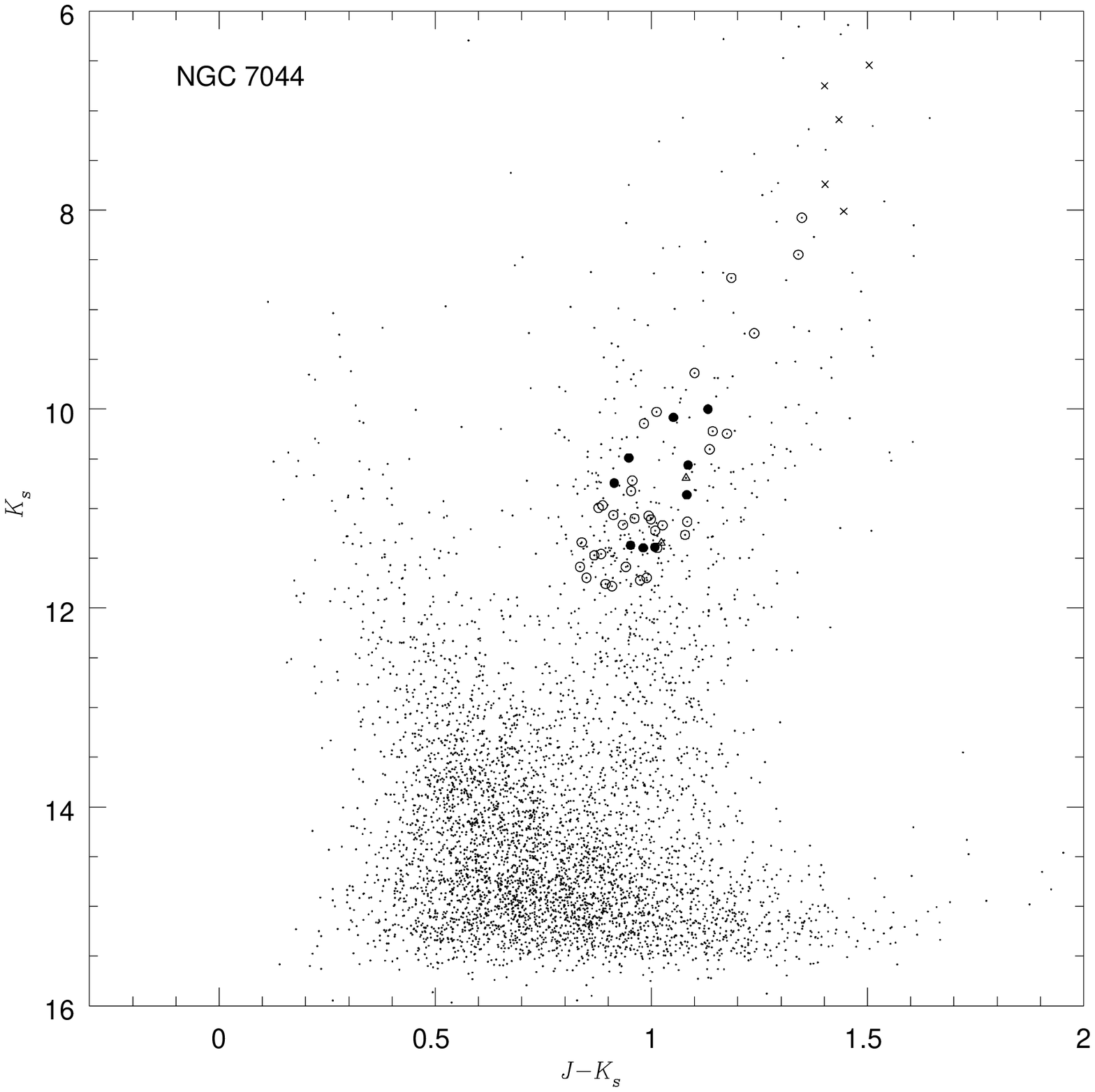}
\caption{The 2MASS ($K_s, J - K_s$) CMD for NGC 7044.  Filled black circles are the accepted cluster stars based upon radial velocity
and positional determinations.  Open symbols are stars rejected from further analysis based
upon radial velocities/falling beyond the visual cluster bounds (circles), TiO bands
(crosses), CN bands (stars), and/or low quality spectra (triangles). \label{ngc7044cmd}}
\end{figure}

As with the calibration clusters, we plotted derived radial velocities versus distance from the 
cluster center in order to select potential cluster members.  
We can also use the 2MASS ($K_s,~J-K_s$) colour-magnitude
diagrams to check if the stars are in the expected RGB sequence of the cluster.
Figures \ref{be81cmd}, 
\ref{be99cmd}, \ref{ic1311cmd}, \ref{king2cmd}, and \ref{ngc7044cmd} show the uncleaned 
2MASS ($K_s,~J-K_s$) CMDs.  The CMDs include all of the point sources in the 2MASS field of view
within a radius of the cluster center determined by the largest distance in the red giant star sample for each
cluster.  As a quick reference, the filled circles highlight the accepted star locations in each 
cluster while the open symbols are those stars rejected from our analysis.  Refer to the
figure captions for  more explanation. 
Figure \ref{rvdistn1}
shows radial velocity vs distance from cluster center.  Again, stars that
have been kept for further analysis are plotted as filled circles and~open symbols are stars
that have been excluded from further analysis.

\begin{figure}
\includegraphics[width=168mm]{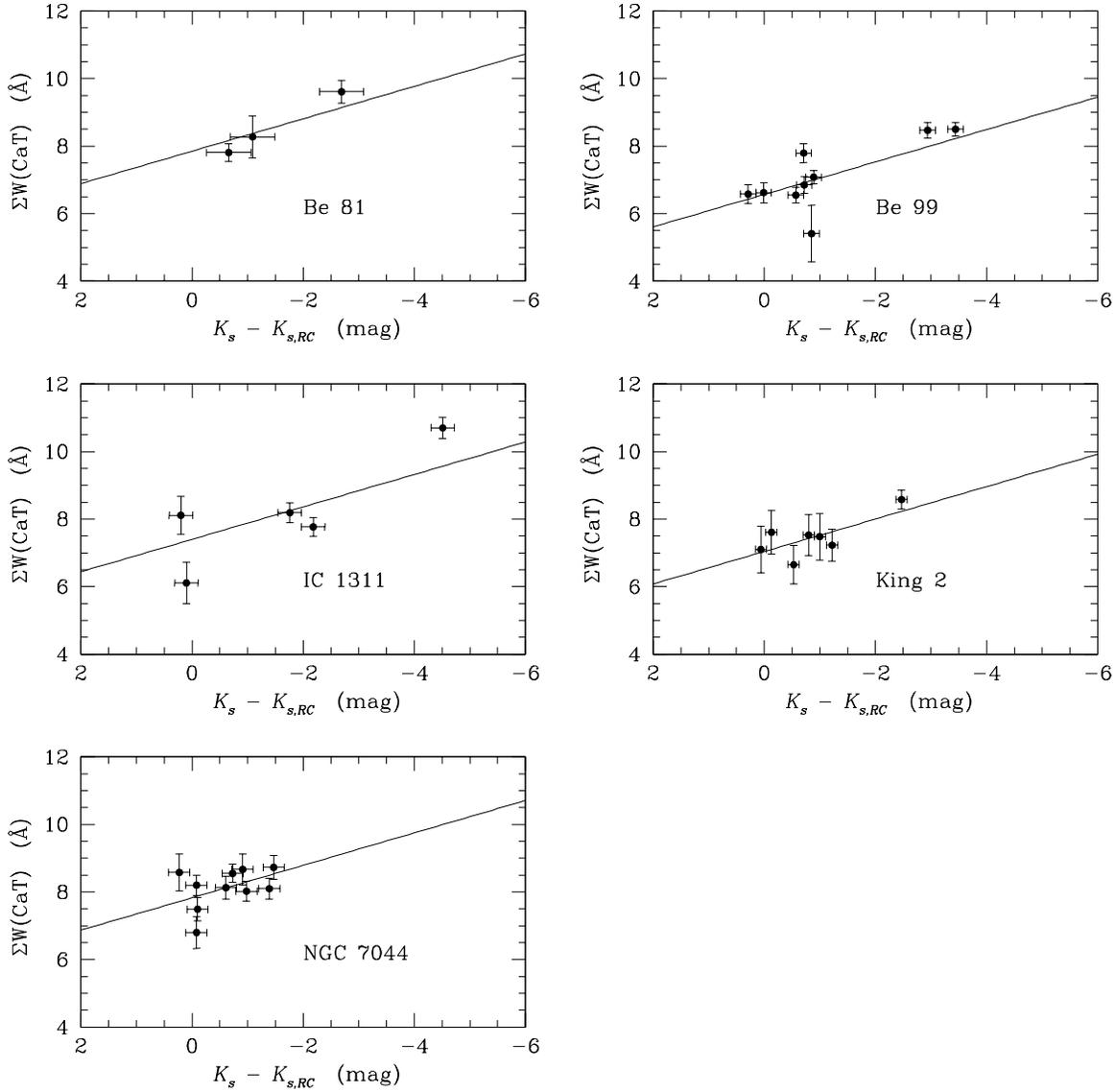}
\caption{The best fit line using $\beta_{K_s}$ = 0.48 as the slope for each cluster.\label{sewmag2}}
\end{figure}

Once our star sample has been defined we can use it to derive cluster parameters.
Figure \ref{sewmag2} is a plot of $\Sigma$W versus $K_s - K_{s,RC}$ for the accepted stars in
each cluster.  As in Figure \ref{sewmag}, the best fit line has a forced slope equal to
our derived $\beta_{K_s}$.  The data points follow the linear relation found previously.  With this
information we can then compute W$^{\prime}$ and use this to derive [Fe/H] for each cluster.  Table
\ref{tab-clus} gives the derived W$^{\prime}$, [Fe/H], and radial velocity values for each of 
the understudied clusters.

\subsection{Berkeley 81 \label{be81-sec}}

\begin{figure}
\includegraphics[width=168mm]{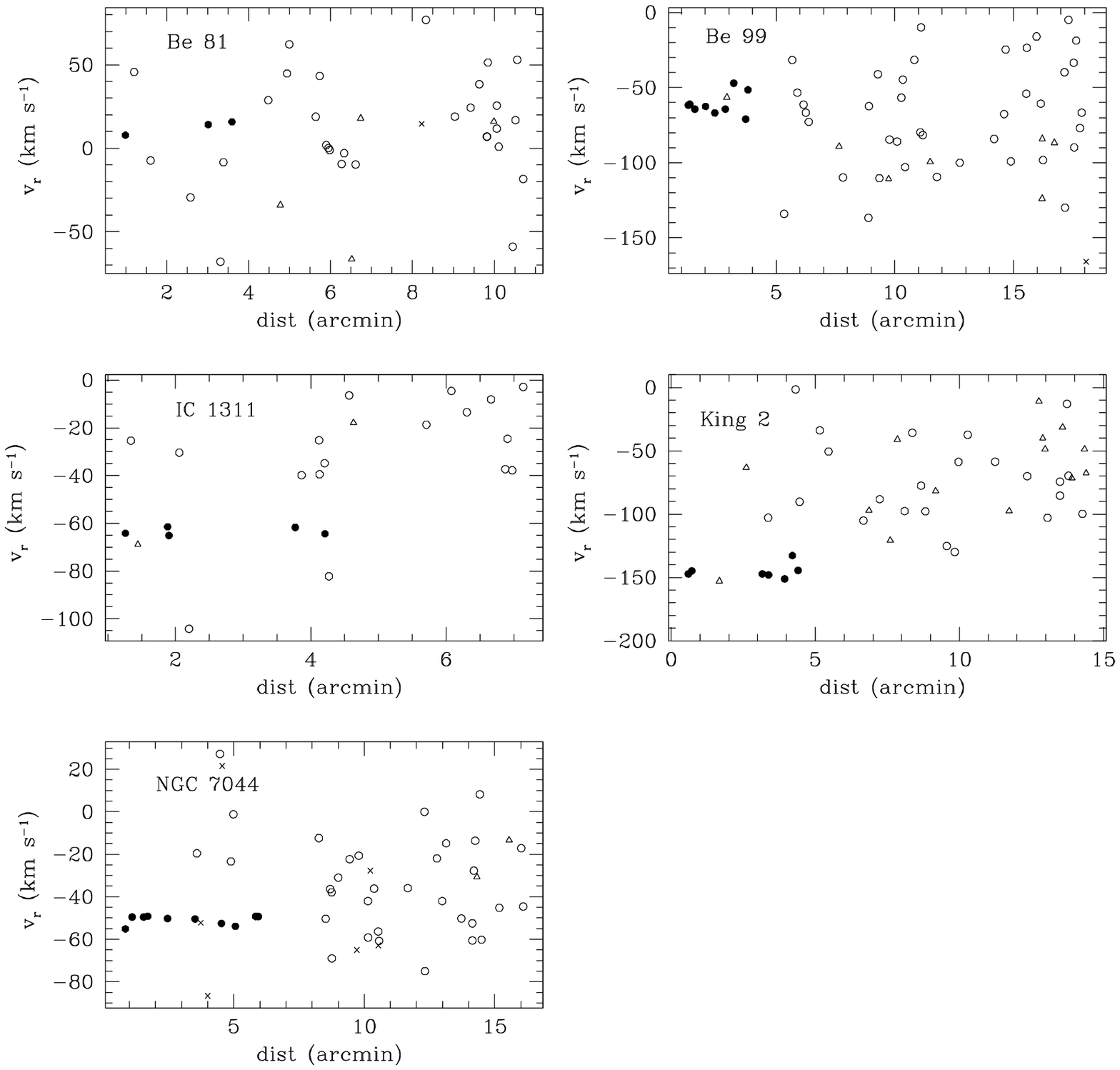}
\caption{Accepted and rejected cluster members in our understudied clusters.
Filled black circles are the accepted cluster stars based upon radial velocity
and positional determinations.  Open symbols are stars rejected from further analysis based
upon radial velocities/falling beyond the visual cluster bounds (circles), TiO bands
(crosses), CN bands (stars), and/or low quality spectra (triangles).
\label{rvdistn1}}
\end{figure}

The cluster sequence for Be~81 was the most difficult of the five new clusters to pick out,
although 
most of the observed spectra for the Be~81 sample were of good quality and only one 
spectrum showed contaminaton by TiO bands. 
With a large sample of stars, we would expect a grouping of stars with similar radial
velocities about the cluster mean in a radial velocity versus radial distance plot; however, 
with the small number of stars in our sample, the 
large spread in radial velocities (see Figure \ref{rvdistn1}) along the cluster line of 
sight made selecting the stars used for further analysis non-trivial.  
Just outside the adopted radius from section
\ref{sample-sec}, the spread of radial velocities is $\approx$ 100 km s$^{-1}$. However, the
8 stars interior to the cutoff also show a large range of radial velocities, and the 
tendency to concentrate around $\approx$ 0 $\pm$ 20 km s$^{-1}$ is not strong, relative to our 
estimated radial velocity uncertainty.  Fig.\ \ref{rvdistn1} shows that 
restricting the sample to the 4 stars within the
cluster radius from \citet{sag98} doesn't appreciably change the situation. 
One possibility is that the radial velocities for this cluster were more poorly
determined than for the other clusters; there is, however, no evidence for this in the spectra 
so the spread in velocity must be real.

Examination
of the CMD suggests that the initial colour selection was too wide.  In light of the
difficulty in isolating the cluster RGB, a wide net was cast in hopes that the cluster
would reveal itself through a well-defined mean radial velocity distinct from the field.
However, we unluckily ended up with stars showing as wide a range in radial velocity
as any stars in the entire field of the spectrograph.  Considering the 
probable age of $\approx$ 1~Gyr from \citet{sag98}, and the richness of the Galactic
field, this should probably have been anticipated and the algorithm for fibre assignments
modified.  Virtually all of the red giants within the cluster radius are RC stars, 
precisely those our algorithm tended to deweight.  The rich background and sparse cluster
upper RGB ensured that the bright stars we preferentially observed were those most likely
to be nonmembers.  Bearing this in mind, we deem it highly unlikely that we failed to
observe {\it any} cluster members, and so we can try to weigh the various probabilities
for membership.

Our initial, very liberal membership cut considered all stars within 1$\sigma$ of the
mean radial velocity of stars within 3$\farcm$6 of the cluster center, v$_r$ = 5.4
$\pm$ 34 km s$^{-1}$.  This left us with six potential members, with an average velocity
v$_r$ = $-$2.3 $\pm$ 17.3 km s$^{-1}$.  This can hardly be considered a restrictive cut,
when the typical velocity dispersion of an open cluster is $\lesssim$ 1 km s$^{-1}$
\citep[e.g.,][]{mer08}, comparable to our 1$\sigma$ random velocity errors.  
All six of these stars are listed in Table \ref{tab-stars}, reflecting the uncertainty
in identifying definite members.

We turned
to photometric information for further guidance regarding the membership probabilities.
The photometric catalogue of \citet{sag98} only covers the southern half of the cluster,
and only includes two of our targets, although all 6 potential members lie within our
adopted cluster radius and the radius given by \citet{lyn87}.  However, the limited 
optical photometry available does appear to rule out the membership of star 2MASS
19012993-0027231, which is far too bright and blue to be a cluster red giant (in the
optical).  This 
star also has a very different radial velocity from the remaining five potential 
members.  That leaves two groups of stars that separate naturally into a group of
three stars at positive radial velocity in a slightly brighter RGB, and a second
group of two stars along a fainter track, with negative radial velocity. 
In either case, we 
apparently have failed to measure any red clump stars, which if nothing else would
have given much better leverage on the radial velocity.  We have a marginal preference 
for the brighter subgroup, with v$_r$ = 13.0 $\pm$ 4.2 km s$^{-1}$, for the following
reasons:  it includes the closest star to the cluster center; the RGB sequence is 
located in a favorable location relative to the red clump in the 2MASS CMD; and
star 2MASS 19013595-00283878, which belongs to the fainter subgroup, was also flagged
as a photometric outlier from the cluster RGB by \citet{sag98}.  With this in mind, 
we take the average metallicity, [Fe/H] = $-$0.15 $\pm$ 0.11, of the brighter subgroup to 
be representative of the cluster.

Be~81 
is our only inner galaxy cluster, lying just interior to the tangent point of
the Sagittarius-Carina spiral arm ($\ell$ = 33.7, $b$ = $-$2.5).  Accordingly, it
is the most highly-reddened cluster in our sample, with E($B-V$) = 1.00 
\citep{dut00}.  \citet{sag98} published the first CCD $BVI$ photometry of Be~81
and derived the accepted distance of 3~kpc and an age of t = 1~Gyr for their
estimate of approximately solar metallicity.  Our derived metallicity,
[Fe/H] = $-$0.15 $\pm$ 0.11, is in agreement with this estimate.
We can check for consistency
with the distance and reddening measurements using the previously determined
behaviour of the magnitude of the red clump, $\langle M_K\rangle$, as a function
of cluster age and metallicity (\citealt{gro02}; \citealt{sal02a}). 
For our measured
clump magnitude $K_{s,RC}$ from Table \ref{clust2} and metallicity from Table
\ref{tab-clus} with the published age and reddening values, the tabulation
of $\langle M_K\rangle$ in \citet{sal02a} yields a distance of 2.6~kpc, slightly
shorter than the 3~kpc measure published by \citet{sag98}.  The 2MASS and optical
CMDs are consistent with this distance, metallicity and reddening for the 
published age of 1~Gyr.  If we had taken as members the fainter, redder red
giants, they would be inconsistent with the position of the RGB from theoretical
isochrones.

The radial velocity of Be~81 is ambiguous:  the three preferred candidate members taken 
here have a mean velocity of 13.0 $\pm$ 4.2 km s$^{-1}$, but the other two possible 
members are nearly identical to each other, with v$_r$ = $-$8.0 $\pm$ 0.7 km s$^{-1}$ 
(however the brighter of the two was excluded from the cluster by \citet{sag98} on the
basis of its $B-V$ colour).
Neither velocity is remarkable compared to the nearby stellar field as observed here.
Additionally, the disk kinematics near the galactic plane can be roughly assessed
by comparison to the velocity profile of the neutral hydrogen, e.g.,
from \citet{har97}.  The H{\sc I} in this part of the galaxy spans a very 
broad range of radial velocities, mostly receding from the Sun.  Both possible
radial velocities of Be~81 are 
well within the range of velocities for gas in the inner galaxy.  Further,
some deviation
from the H{\sc I} rotation curve is expected for old open clusters,
which in general lag the thin disk rotation and show a significant number
of eccentric galactic orbits \citep{sco95}.

\subsection{Berkeley 99}

Like Be~81, the spectra for the Be~99 star sample were of good quality with only one star
showing TiO bands.  The radial velocity versus radial distance plot (Figure \ref{rvdistn1})
shows a tight grouping of stars around v$_r$ $\approx$ $-$60 km s$^{-1}$ which we adopted to be
cluster stars.  The outer limit of the cluster was set at r $\approx$ 5$\arcmin$ based 
on the 2MASS CMD, larger than the radius in \cite{sag98} (the most distant star in the
final sample was $\approx$ 4$\arcmin$ away).  See Table \ref{tab-stars} for information on
the 9 stars in the final sample.

Be~99 and King~2 are the two clusters in our sample that are significantly 
outside the solar circle.  Be~99 is both sparser and more distant than Be~81 \citep{sag98},
but also less reddened with E($B-V$) $\approx$ 0.3--0.45 \citep{dut00,sfd98}.  Although Be~99 is higher off 
the Galactic plane than any of our other clusters ($b$ = $+$10), its measured
radial velocity is in excellent agreement with the H{\sc I} rotation for its
longitude ($\ell$ = 116). 

Be~99 has the lowest
metallicity of any cluster in our sample, but it is not unusual for its
galactocentric radius \citep{twa97,friel02}.  Its metallicity of [Fe/H] = $-$0.58 $\pm$ 0.10
mandates a re-examination of its distance, age, and reddening.  Comparison of
theoretical isochrones to the 2MASS ($K_s,~J-K_s$) CMD and to the data of \citet{sag98}
reveals that the reddening E(B$-$V) must be toward the high end of the published values:
E(B$-$V) $\approx$ 0.43.
The lower metallicity and higher reddening combine to give consistency with the observed
colour of the main sequence and RC.  A downward revision of the age to $\approx$ 2.5~Gyr
and retention of the accepted distance of 5~kpc (distance modulus (m$-$M)$_0$ = 13.49)
reproduce the magnitude of 
the red clump in both $K_s$ and I bands and is consistent with the main-sequence turnoff.

\subsection{IC 1311}

The star sample for IC~1311 contained only two low quality spectra with no stars showing
contamination of TiO or CN bands.  Inspection of the radial velocity versus radial distance
plot (Figure \ref{rvdistn1}) showed two groupings of stars, one around v$_r$ $\approx$ $-$30 km
s$^{-1}$ and the other at v$_r$ $\approx$ $-$65 km s$^{-1}$.  
Attempts to identify trends in W$^{\prime}$ versus radial distance were of
little help in determining which targets belonged to the cluster and which were field stars. 
We plotted both groupings on a 2MASS ($K_s, J-K_s$) colour-magnitude diagram (see Figure 
\ref{ic1311cmd}) in order to determine which locus each occupied.  The grouping with 
v$_r$ $\approx$ $-$65 km s$^{-1}$ traced the red giant branch more closely and was adopted as the
cluster sample.  The set of stars at this velocity is more centrally concentrated than the
higher velocity set, and has a smaller range of velocities.  Our final sample of 5 stars
is listed in Table \ref{tab-stars}, and extends out to $\approx$ 4$\farcm$5 radius, 
larger than the initial radius used to define the red clump magnitude but consistent with
the visual appearance of the cluster.

The first CCD $UBVR$ photometry of this relatively rich cluster was published by
\citet{del94}, who derived a distance of 6~kpc and an age of 1.6~Gyr for
[Fe/H] = 0 and E($B-V$) = 0.28 mag.  The only other metallicity estimate of which
we are aware is from the compilation of \citet{tad01}, who estimated [Fe/H] = $-$0.23
(no uncertainty is given) based on the classical ($U-B$) colour excess of the cluster.
Our derived metallicity, [Fe/H] = $-$0.30 $\pm$ 0.16,
is not significantly different from either value but appears to confirm that IC~1311
has slightly subsolar metallicity.
Adopting the age, distance, and reddening from \citet{del94} would produce a red clump magnitude
$K_{s,RC}$ = 12.48, nearly half a magnitude brighter than observed.  To adopt the
reddening value advocated by \citet{dut00}, E($B-V$) = 0.45, would worsen
the discrepancy,
because consistency with the visible-light CMD would require a corresponding
shift in distance modulus $\Delta$(m$-$M)$_0$ $\approx$ $-$0.5 mag, producing an
even brighter $K_{s,RC}$.  Visual inspection of the \citet{del94} ($V, B-V$)
CMD combined with our observed $K_{s,RC}$ suggests that consistency can be
achieved for a larger distance, d $\approx$ 6.6~kpc, and a reddening in the range 
0.3 $<$ E($B-V$) $<$ 0.4 mag.  This requires a revision of the cluster age
to t $\approx$ 1.1~Gyr.  An age near 1~Gyr is also consistent with the relatively 
high specific frequency of red clump stars in the cluster \citep{sal02a}.
Note that the age listed in \citet{dia02} and in the online WEBDA database as of
early 2008, log(t) = 8.625, is obviously far too young to be consistent with the \cite{del94}
or 2MASS CMDs.  The revised distance puts IC~1311 clearly beyond
the solar circle, but still within 10~kpc of the Galactic center for R$_0$ = 8~kpc.
The measured heliocentric radial velocity of $-$63 km s$^{-1}$ is in excellent agreement
with the velocity of peak Galactic H{\sc I} (thin disk) in this direction.

\subsection{King 2}

The stellar sample for King~2 had a larger quantity of poor-quality spectra than any of
the other clusters in our sample.  Fortunately all but two of the lower quality spectra stars 
fell outside of the adopted bounds of the cluster (r $\approx$ 5$\arcmin$). 
The cluster radial velocity sequence was simple to pick
out on the radial velocity versus radial distance plot (Figure \ref{rvdistn1}) at around 
v$_r$ $\approx$ $-$145 km s$^{-1}$.  The final King~2 star sample consists of 7 stars (see Table
\ref{tab-stars}).

King~2 is less studied than IC~1311 despite being of similar richness, distance,
and reddening \citep[e.g.,][]{lyn87,dia02}.  It certainly appears far less conspicuous on the sky,
presumably because its greater age has truncated the main-sequence at much fainter magnitudes.
The CMD study by \citet{kal89} resulted in a range of plausible ages
and distances for different assumed reddenings and metallicities, while \citet{apa90}
derived an age of 6~Gyr and a distance of 5.7 kpc for solar metallicity.  \citet{tad01}
derives 
the ($U-B$) colour excess from the literature data and makes an estimate of [Fe/H] = $-$0.32
(no uncertainty given).  Our CaT spectroscopy yields [Fe/H] = $-$0.42 $\pm$ 0.09; for this
metallicity, \citet{kal89} claims a most probable distance
of 6.9~kpc and an age of $\approx$ 5~Gyr \citep[see also][]{sal04}. 
The metallicity is significantly subsolar, 
inconsistent with the finding by \citet{apa90}.
We find that a distance of 6.5~kpc and
a slightly younger age, t $\approx$ 4~Gyr, better fits the optical CMD and 2MASS $K_{s,RC}$ if
the reddening E($B-V$) = 0.31 from \cite{dut00} is adopted.  The new distance puts it 
at R$_{GC}$ $\approx$ 13~kpc, where its metallicity falls close to the trend of the
galactic abundance gradients derived in \citet{friel02}.  However, the observed radial
velocity, v$_r$ = $-$144.2 $\pm$ 5.9 km s$^{-1}$, differs by $\approx$100 km s$^{-1}$ from 
the rotation
velocity of the gas disk at its longitude $\ell$ = 123$^{\circ}$.  This surprising deviation is
discussed in detail in section \ref{sec-mon} below.

\subsection{NGC 7044}

The star sample for NGC~7044 contained a large number of stars (6) showing the effects of TiO
bands.  These were the six brightest ($K_s~<~8.02$) and reddest ($J-K_s~>~1.39$) stars in
the sample.  Only three of these stars fell within the adopted cluster radius (r $\approx$
6$\arcmin$) while only one showed a similar radial velocity to the other adopted cluster stars
(v$_r$ $\approx$ $-$50 km s$^{-1}$).  The adopted cluster stars showed a tight radial velocity
grouping (Figure \ref{rvdistn1}) with very little spread making the selection of members
relatively straightforward.  Table \ref{tab-stars} lists the pertinent information
for the ten stars in the NGC~7044 sample.

NGC 7044 is the most well-studied of the five clusters presented here, with three CCD
photometric studies in the literature: (\citealt{kal89}, \citealt{apa93}, and \citealt{sag98}).
It has also been included in the overview of open cluster age measurements from cluster
morphology in \cite{sal04}; the age measurements produced a range from 1.0--2.5~Gyr, and
distances from 3--4~kpc.  Because
the metallicity [Fe/H] = $-$0.16 $\pm$ 0.09 is consistent with the assumption of solar 
metallicity, no revision to the age is mandated by the new determination.  However, 
consistency with $K_{s,RC}$ rules out a distance as high as 4~kpc.
Adopting the reddening value E($B-V$) = 0.67 from \citet{dut00}, the
1~Gyr age of \citet{sal04} yields a consistent $K_{s,RC}$ for d = 3.2~kpc, while the
1.6~Gyr age from \citet{sag98} requires d $\approx$ 3.6~kpc.  The latter pair of
parameters provides greatest consistency with the CMD data from \citet{apa93}
and \citet{sag98}.  Note that the 1.6~Gyr age is consistent with the conclusions
of \citet{del94}, who re-examined the data of \citet{apa93} in their study of IC~1311
and found that NGC~7044 is a few hundred Myr older than IC~1311.
The new distance places NGC~7044 just beyond the solar circle.  It has a slightly
higher than typical metallicity, but lies within the scatter of the general galactic
abundance gradient \citep{twa97,friel02}.  The radial velocity v$_r$ = $-$50.6 $\pm$ 2.2 
km s$^{-1}$ is in excellent agreement with the galactic rotation velocity of both
H{\sc I} and older disk stars \citep{har97, sco95, car07a}.

\section{Is King~2 Associated with the Monoceros Stream?}
\label{sec-mon}

The velocity distribution for the King~2 field is shown in 
Figure~\ref{rvdistn1}.
The cluster clearly stands out, far from the
general outer disk field along its sightline.  We examined the predicted
velocity distribution for stars at $\ell$ = 123$^{\circ}$, b = $-$5$^{\circ}$
in the Besan\c{c}on Galactic model of \citet{rob03}; for heliocentric
distances greater than 3~kpc, fewer than 1\% of all disk stars in the model
have both v$_r$ $\leq$ $-$130 km~s$^{-1}$ and [Fe/H] $\geq$ $-$0.5, consistent
with our easy identification of cluster members.
The discrepant heliocentric radial velocity of King~2 compared
to other disk populations leads us to look for possible explanations.

If King~2 is moving in pure circular motion, its observed radial velocity
translates to an orbital speed of just 80~km s$^{-1}$, far too low to be a normal
disk member.  Conversely, disk membership and orbital speeds of 150--200~km s$^{-1}$
would imply large proper motions for the cluster as a large fraction of the 
motion would be perpendicular to the line of sight.  Unfortunately,
King~2 is the only cluster in our sample that does not have an estimated proper
motion from \citet{dia06} or \citet{kha03}, so that three-dimensional 
velocity information is unavailable to help understand King~2's place
in the Galaxy.

While many old open clusters follow eccentric galactic orbits \citep{sco95},
King~2 is further out of synch with disk rotation than even Be~17, an anticentre
cluster noted for its unusual radial velocity.  
Based on its relatively young age and location in the outer Galaxy,
bulge membership
would appear to be a poor explanation for the kinematics of King~2.
However a disk origin within the solar circle may be plausible; if
the entirety of King~2's motion is directed along its galactocentric
radius vector, then it is moving radially {\it outward} at nearly
50~km s$^{-1}$.  If on the other hand it is orbiting at typical 
thick disk speeds of $\approx$ 180~km s$^{-1}$, then it must be infalling
at speeds in excess of 60~km s$^{-1}$.  Whatever the true situation,
it appears likely that King~2 is on a high-eccentricity, low
angular momentum path around the Galaxy.

Assessing the degree to which King~2's motion is anomalous
is hindered by the complexity of kinematics in this
part of the galaxy.
The Perseus spiral arm shows anomalous kinematics
possibly related to its participation in a density wave, but lies 3--4
times closer to us than King~2 \citep{xu06}; On the far side of King~2,
the high-velocity cloud Complex~H (HVC130$+$01$-$161) overlaps 
in radial velocity
with the cluster, but is probably several times more distant \citep{wak01}.
Further complicating the picture, King~2 lies near on the sky to 
where several authors have found extensions of the structure known as
the Monoceros stream \citep{new02, iba03, yan03}.  Summaries of kinematic
data pertaining to the Mon stream can be found in, e.g., \citet{pen05,you08}.

Could King~2 be a member of the Monoceros stream?  Our favoured distance
of $\approx$ 6.5~kpc places the cluster at a galactocentric distance of
$\approx$ 13~kpc, consistent with but slightly interior to the Mon stream
distance favoured by \citet{iba03}, and near the inner limit of detections
reported from SDSS data in \citet{new02}.  Both \citet{new02} and 
\citet{roc03} detected Mon stream components among field stars at the 
longitude of King~2, but these earlier detections were at much higher positive
and negative galactic latitude than King~2's relatively modest
$b$ = $-$5$^{\circ}$.  \citet{roc04} measured radial velocities for stars
over the longitude range from 103--163$^{\circ}$, 
and their reported velocities correspond exactly to King~2's v$_r$ = $-$144~km s$^{-1}$;
however, their measurements apparently pertain to stars averaging
nearly 3 times farther away than King~2 (in the Triangulum-Andromeda
structure).  This, and the
discrepant galactic latitudes, suggests that either the stream
is extremely extended or the agreement is fortuitous.
However, the distances found by \citet{roc04} are strongly metallicity
dependent, and \citet{roc03} already noted the presence of a structure
at closer distances. Further, the models of \citet{pen05} suggest that the
majority of stars in the Monoceros stream should lie more nearby
than the Tri-And structure, with similar radial velocities to the more
distant stars.

If King~2 is indeed a member of the Monoceros stream, we might expect to
see other stream members in the surrounding field.  \citet{roc03} show
that $\approx$ 10\% of the red giants in this general direction are distinct
from the foreground of ordinary thick disk stars in their plots of 
relative distance vs.\ longitude.  Of the 36 non-members we find in the King~2 field,
two have radial velocities within the reported velocity dispersion of 
the stream measured by \citet{roc04}, $\sigma_{v_r}$ = 20~km s$^{-1}$.  
Fig.~\ref{rvdistn1} shows that these two stars have similar {\it reduced}
CaT equivalent widths to the King~2 giants; however, since their distance
isn't known, no accurate metallicity can be assigned to them.  Their
equivalent widths are consistent with being at the same distance and metallicity
as King~2, but also with a distance 3 times larger and a metallicity 0.4~dex
lower.

One interpretation of the Mon stream is as the tidally-disrupted 
remains of a satellite dwarf galaxy consumed in a minor merger with the 
Milky Way \citep[e.g.,][]{hel03,mar05}.  However, star clusters are uncommon in
the known dwarf spheroidal (dSph) companions of the Milky Way:
Fornax and Sagittarius
host a few globular clusters each, but open clusters in dwarf spheroidals are
unknown.  Presumably this is not an absolute exclusion of open clusters in 
dSphs, but simply reflects the fact that star-formation rates have been low
over longer timescales than the typical disruption time of a low-density cluster.
Ownership of several globular clusters and open clusters has been attributed
to the Monoceros stream and associated structures \citep[e.g.,][]{fri04,pen05},
although most of those have been towards the anticentre or third galactic 
quadrant and at larger distances.  Detailed modelling, such as in \citet{pen05}
has cast doubt on a number of these associations.

King~2's association with the Monoceros stream can remain only speculative
in the absence of proper motion data, but it is an intriguing explanation for
its anomalous kinematics.  If the Monoceros stream does in fact claim King~2
as a member, then this could potentially shed light on the nature of the 
stream.  If the Mon stream is really due to the dissolution of a 10$^8$--10$^9$
M$_{\sun}$ dwarf galaxy in the plane of the Milky Way,
then there may be multiple wraps of the tidal debris around the Galactic disk; 
the distance of King~2 would place it in the innermost of the rings, most recently
stripped from its host.  Its 
metallicity and galactic latitude might then help resolve some of the ambiguities in
current models \citep[e.g.,][]{pen05}.

We also note that while King~2 has a higher metallicity than any dwarf galaxy
except Sagittarius,
it has a perfectly ordinary metallicity 
for an outer disk cluster \citep{twa97,friel02}.  Some controversy also persists
as to the external, dwarf galaxy nature of both the Monoceros stream and the
Canis Major dwarf (or overdensity) that is often taken to be the
progenitor of the stream \citep[e.g.,][]{car08}.
\citet{you08} present detailed
models of the high orbital eccentricity flyby of a large dwarf spheroidal, and
show that the structure excited in the Milky Way disk strongly resembles a stream
or ring similar to the Monoceros stream.  King~2 may simply be an ordinary Milky Way
old open cluster,  participating in one of
the ripples left in the wake of a satellite close encounter as described by
\citet{you08}; this has the advantage of naturally accounting for King~2's 
metallicity.  

\section{Discussion and Conclusions}

We have obtained near infrared spectra ($\lambda$ = 8416 - 8986 \AA) of 133 red giant stars from
two Galactic globular clusters and ten Galactic open clusters and combined this sample with that of \citet{cole04}.  
We used the calcium
triplet lines to 
derive metallicities following the method pioneered by \citet{arm91} and calibrated with
Galactic globular clusters by 
\citet{rutl97b}.  Our cluster sample spanned 2.2 dex in metallicity and 11 Gyr in age,
comparable to the widest ranges in the literature to date \citep{car07}.

All prior studies of this technique that used both globular and open clusters
\citep[e.g.,][]{cole04,car07}
have relied upon multiple literature sources for
$VI$ photometry. The use of numerous sources for photometry can lead to a non-uniform sample
and could introduce non-linear effects in their calibration.  Using 2MASS $K_s$ photometry to calibrate our metallicity scale 
has the benefit of a uniform photometry sample for all stars down
to $K_s$ $\approx$ 14.  This is very important because it allows a unique opportunity to easily
derive the metallicity for any Galactic cluster with only one new medium resolution spectra 
observation.  If differential reddening plays a role, as it does
in certain bulge globulars, use of $K_s$ magnitudes will dramatically reduce its impact,
perhaps allowing future studies to distinguish claims for internal abundance spreads from
variable reddening.

Like others before us
\citep[e.g.,][]{car01,cole04,car07}, our data show 
no signs of a non-linear relationship between CaT equivalent widths and metallicity. 
Therefore we can say that the slope of the best fit line in the $\Sigma$W versus
$K_s-K_{s,HB}$
plane is independent of metallicity over the small range in magnitudes used in the RGB.
We are able to use our calibration in estimating the [Fe/H] values for five understudied
Galactic open clusters (Be~81, Be~99, IC~1311, King~2, and NGC~7044).  All of our metallicity 
estimates are statistically similar to estimates obtained by other methods, supporting the
use of the calcium triplet plus 2MASS near-infrared magnitudes to derive [Fe/H] for any cluster.

Calibrating the CaT - metallicity relation using 2MASS data has the added benefit in which any
new observations in the $K$ band can be calibrated to the 2MASS scale \citep{carp01}.  This
allows our technique to be extended to extragalactic sources (i.e., the Magellanic clouds) and
still retain uniformity.  Future work can be done to use the tip of the RGB instead of the
horizontal branch/red clump to define W$^{\prime}$ in terms of ($\mathfrak{M}-\mathfrak{M}_{TRGB}$).
This would allow us to use the 2MASS photometry to estimate the metallicity of LMC/SMC clusters
where 2MASS is not deep enough to reach the horizontal branch/red clump, obviating the need
for new photometric data.

In order to derive the greatest benefit from this technique, we would like 
to use a uniform and consistent high-dispersion metallicity scale for
both globular and open clusters.  There are currently no studies for which a large sample of
both families of clusters have had high-dispersion metallicities derived using atmospheric
modelling.  Because of this our calibration is referred back to different reference scales in 
[Fe/H] (\citealt{cg97}; \citealt{friel02}).  Our reported cluster reduced equivalent
widths can be converted to any desired metallicity scale simply by rederiving the coefficients
of the equation relating W$^{\prime}$ and [Fe/H].  Considering a very short list of 
alternative calibrator metallicities shows that the absolute calibration strongly changes
at the high end, with clusters like Be~81 and NGC~7044 moving upward to solar 
metallicity if the latest high-dispersion abundances for clusters like NGC~6939 and NGC~7142
are chosen as the reference.

We report the first radial velocities and spectroscopic metallicities for five old
open clusters chosen solely on the basis of their red giant content.   In most cases
the derived abundances are consistent with optical CMD morphology, the $K_s$-magnitude
of the red clump, and published distance, age, and reddening estimates.  The cluster
Berkeley~81 was difficult to pick out from the field on the basis of radial velocity
or CMD,
re-emphasizing the limitations on such studies
imposed by the generally sparse open cluster red giant numbers \citep{can70,twa97}.
However, unambiguous cluster sequences were defined in the other four cases.  Berkeley~99
was found to be the most metal-poor cluster in the understudied sample, with [Fe/H]
= $-$0.58 $\pm$ 0.10; the interplay between distance, reddening, and metallicity
suggests that the age is slightly lower than reported in \citet{sag98}.  IC~1311
must be more distant than previously reported, at $\approx$ 6.6~kpc; it also must be
significantly older than the 650~Myr reported in \citet{car94}.  All clusters except
King~2 were found to be unexceptional in their radial velocities \citep{sco95}, with 
the caveat that the Be~81 velocity remains ambiguous because there was no clear way
to distinguish between member and nonmember giants in the absence of a larger 
spectroscopic sample or more extensive optical photometry.

The cluster King~2 is strongly out of galactic rotation, lagging the disk by 60--100~km s$^{-1}$.
Its heliocentric radial velocity v$_r$ = $-$144~km s$^{-1}$ corresponds precisely to the velocities
of stars in the Monoceros stream at its longitude of $\ell$ = 123$^{\circ}$, although there
may be discrepancies in distance and galactic latitude \citep{roc04,pen05}.
Its metallicity [Fe/H] = $-$0.42 $\pm$ 0.09 is high for a dwarf spheroidal origin, but 
very ordinary for old Milky Way open clusters.  Apart from an extragalactic origin,
King~2's velocity may be due to the ejection of an inner disk cluster into a high-eccentricity
orbit due to interactions, e.g., with the Galactic bar, or to its participation in ripples
induced in the wake of a satellite galaxy flyby \citep{you08}.  The possibilities cannot
be disentangled in the absence of proper motion information.

\section*{Acknowledgments}Travel support for AAC was provided by the Netherlands Research 
School for Astronomy (NOVA).  We would like to thank our referee, Dr. Bruce Twarog, for his 
comments and suggestions that have greatly improved the manuscript.  
AAC would like to thank ING support astronomer
Ian Skillen and night assistant Roberto Martinez for their able assistance
during the observing run.  Thanks to Professor Linda Sparke for helpful information
about the kinematics of the disk and to Professor Evan Skillman for helpful comments 
about the text.
This publication makes use of data products from the Two Micron All Sky Survey,
which is a joint project of the University of Massachusetts and IPAC/Caltech,
funded by NASA and the NSF.  This research has made use of the WEBDA database,
operated at the Institute for Astronomy of the University of Vienna.

\clearpage

\begin{table}
\begin{center}
\caption{Previously Measured Clusters\label{clust1}}
\begin{tabular}{lcccccc}
\hline\hline
Cluster&$\alpha$ (J2000) & $\delta$ (J2000) &[Fe/H]
& Age (Gyr) & $K_{s,RC}$ & References\\
\hline
M15         &$21^{\mbox{h}}29^{\mbox{m}}58\fs3^a$ &$+12\degr 10\arcmin 04\farcs6^a$ &$-2.12\pm 0.01$ &$11.7$ &$14.83\pm 0.35$ &1,2\\
M71         &$19^{\mbox{h}}53^{\mbox{m}}46\fs7^a$ &$+18\degr 46\arcmin 45\farcs7^a$ &$-0.70\pm 0.03$ &$10.2$ &$11.83\pm 0.07$ &1,2\\
NGC 6791    &$19^{\mbox{h}}20^{\mbox{m}}50\fs7^a$ &$+37\degr 45\arcmin 38\farcs4^a$ &$+0.11\pm 0.10$ &$10.2$ &$11.48\pm 0.08$ &3,4\\
NGC 6819    &$19^{\mbox{h}}41^{\mbox{m}}14\fs5^a$ &$+40\degr 11\arcmin 58\farcs0^a$ &$-0.11\pm 0.06$ &$2.9$  &$10.27\pm 0.11$ &3,4\\
NGC 6939    &$20^{\mbox{h}}31^{\mbox{m}}40\fs7^a$ &$+60\degr 39\arcmin 34\farcs5^a$ &$-0.19\pm 0.09$ &$2.1$  &$9.89\pm 0.17 $ &3,4\\
NGC 7142    &$21^{\mbox{h}}45^{\mbox{m}}03\fs8^a$ &$+65\degr 45\arcmin 33\farcs3^a$ &$-0.10\pm 0.10$ &$4.0$  &$10.36\pm 0.30$ &3,4\\
NGC 7789    &$23^{\mbox{h}}57^{\mbox{m}}29\fs4^a$ &$+56\degr 43\arcmin 38\farcs4^a$ &$-0.24\pm 0.09$ &$1.8$  &$10.05\pm 0.18$ &3,4\\
\hline
\citet{cole04} Clusters&&&&&&\\
\hline
Berkeley 20 &$05^{\mbox{h}}32^{\mbox{m}}34^s$     &$+00\degr 10\arcmin            $ &$-0.61\pm 0.14$ &$4.0$  &$13.20\pm 0.15$ &3,4\\
Berkeley 39 &$07^{\mbox{h}}46^{\mbox{m}}45^s$     &$-04\degr 41\arcmin            $ &$-0.26\pm 0.09$ &$7.0$  &$11.54\pm 0.09$ &3,4\\
M67         &$08^{\mbox{h}}51^{\mbox{m}}25^s$     &$+11\degr 48\arcmin            $ &$-0.15\pm 0.05$ &$4.3$  &$ 7.95\pm 0.02$ &3,4\\
Melotte 66  &$07^{\mbox{h}}26^{\mbox{m}}28^s$     &$-47\degr 41\arcmin            $ &$-0.47\pm 0.09$ &$5.3$  &$11.74\pm 0.13$ &3,4\\
NGC 104     &$00^{\mbox{h}}26^{\mbox{m}}33^s$     &$-71\degr 51\arcmin            $ &$-0.70\pm 0.07$ &$10.9$ &$11.94\pm 0.12$ &1,4\\
NGC 1851    &$05^{\mbox{h}}14^{\mbox{m}}15^s$     &$-40\degr 04\arcmin            $ &$-0.98\pm 0.06$ &$9.2$  &$14.24\pm 0.18$ &2  \\
NGC 1904    &$05^{\mbox{h}}24^{\mbox{m}}12^s$     &$-24\degr 31\arcmin            $ &$-1.37\pm 0.01$ &$11.7$ &$13.90\pm 0.30$ &1,2\\
NGC 2141    &$06^{\mbox{h}}03^{\mbox{m}}00^s$     &$+10\degr 30\arcmin            $ &$-0.33\pm 0.10$ &$2.5$  &$11.53\pm 0.10$ &3,4\\
NGC 2298    &$06^{\mbox{h}}48^{\mbox{m}}59^s$     &$-36\degr 00\arcmin            $ &$-1.74\pm 0.06$ &$12.6$ &$14.70\pm 0.20$ &1,2\\
NGC 4590    &$12^{\mbox{h}}39^{\mbox{m}}28^s$     &$-26\degr 45\arcmin            $ &$-1.99\pm 0.10$ &$11.2$ &$14.50\pm 0.10$ &1,2\\
\hline
\end{tabular} 
\end{center}
$^a$ Center of AF2/WYFFOS field.\\
References to metallicity and age values: (1) \citet{cg97}, (2) \citet{sal02}, (3) \citet{friel02}, (4) \citet{sal04}.
\end{table}

\begin{table}
\begin{center}
\caption{Clusters Without Previous Metallicity Determinations\label{clust2}}
\begin{tabular}{lcccccccc}
\hline\hline
Cluster&$\alpha$ (J2000)$^a$ & $\delta$ (J2000)$^a$
&$\ell$ (deg)&$b$ (deg)&dist.\ (kpc) &Age (Gyr) &$K_{s,RC}$ &Reference\\
\hline
Berkeley 81 & $19^{\mbox{h}}01^{\mbox{m}}40\fs2$ & $-00\degr 27\arcmin 26\farcs3$ & 33.7 & -2.5 & 3.0 &$1.0$ & $11.27\pm 0.40$ &1\\
Berkeley 99 & $23^{\mbox{h}}21^{\mbox{m}}14\fs2$ & $+71\degr 46\arcmin 50\farcs7$ & 116.0 & +10.1 & 4.9 &$3.2$ & $12.05\pm 0.14$ &1\\
IC 1311     & $20^{\mbox{h}}10^{\mbox{m}}49\fs4$ & $+41\degr 10\arcmin 46\farcs0$ & 77.7 & +4.2 & 5.3 & $1.1$ & $12.96\pm 0.21$ &2\\
King 2      & $00^{\mbox{h}}50^{\mbox{m}}57\fs6$ & $+58\degr 11\arcmin 20\farcs8$ & 122.9 & -4.7 & 6.5 & $5.0$ & $12.51\pm 0.10$ &3\\
NGC 7044    & $21^{\mbox{h}}13^{\mbox{m}}05\fs4$ & $+42\degr 29\arcmin 47\farcs0$ & 85.9 & -4.2 & 3.1 & $1.6$ & $11.47\pm 0.19$ &1\\
\hline
\end{tabular} 
\end{center}
$^a$ Center of AF2/WYFFOS field.\\
References to age values: (1) \citet{sag98}, (2) \citet{del94}, (3) \cite{kal89}.
\end{table}

\begin{table}
\begin{center}
\caption{Defined Line and Continuum Bandpasses (\AA)\label{bpass}}
\begin{tabular}{lcc}
\hline\hline
Blue Continuum&Line&Red Continuum\\
\hline
8474-8489    &8490-8506    &8521-8531 \\
8521-8531    &8532-8552    &8555-8595\\
8626-8650    &8653-8671    &8695-8725\\
\hline
\end{tabular}
\end{center}
\end{table}

\clearpage

\begin{table}
\begin{center}
\caption{Sample of Cluster Giants and Derived Parameters$^a$ \label{tab-stars}}
\begin{tabular}{lccccccccl}
\hline\hline
2MASS ID & $K_s$ & $\sigma _{K_{s}}$ & v$_r$ & $\sigma _{v_{r}}$ &
$\Sigma$W(CaT) & $\sigma _W$ & Notes \\
& (mag) & (mag) & (km s$^{-1}$) & (km s $^{-1}$)& (\AA) & (\AA) & \\
\hline
\multicolumn{8}{l}{M15}\\
21295801+1214260 & 12.45 & 0.02 & -108.79 & 2.04 & 3.39 & 0.35 & \\
21295560+1212422 & 10.61 & 0.02 & -112.75 & 1.66 & 3.83 & 0.17 & \\
21294979+1212298 & 12.96 & 0.03 & -108.12 & 2.25 & 2.94 & 0.37 & \\
21294607+1211315 & 11.96 & 0.02 & -108.56 & 1.76 & 3.86 & 0.32 & \\
21294351+1210033 & 11.89 & 0.02 & -104.09 & 1.60 & 3.45 & 0.28 & \\
21294993+1208052 & 10.73 & 0.02 & -99.23 & 1.38 & 4.08 & 0.24 & \\
21300038+1207363 & 10.85 & 0.02 & -100.52 & 0.94 & 3.80 & 0.19 & RV template star \\
21300637+1206592 & 10.70 & 0.02 & -108.89 & 0.03 & 3.91 & 0.26 & RV template star \\
21300750+1208136 & 11.21 & 0.02 & -111.78 & 1.16 & 4.01 & 0.20 & \\
21301049+1210061 & 10.44 & 0.02 & -98.90 & 1.13 & 4.82 & 0.21 & \\
21301522+1211345 & 12.79 & 0.03 & -111.59 & 1.79 & 3.11 & 0.44 & \\
21300978+1212544 & 11.40 & 0.02 & -99.89 & 1.17 & 4.14 & 0.27 & \\
21300316+1213286 & 11.51 & 0.02 & -109.68 & 0.01 & 3.63 & 0.24 & RV template star \\
\hline
\multicolumn{8}{l}{M71}\\
19534463+1851392 & 11.00 & 0.02 & -23.44 & 1.22 & 6.65 & 0.19 & \\
19533399+1849186 & 11.92 & 0.02 & -22.98 & 1.46 & 6.49 & 0.25 & \\
19533757+1847286 & 8.42 & 0.01 & -29.65 & 1.32 & 8.23 & 0.18 & \\
19533470+1846213 & 10.79 & 0.02 & -26.82 & 1.09 & 6.86 & 0.31 & \\
19532342+1845023 & 11.87 & 0.02 & -20.34 & 1.73 & 6.53 & 0.22 & \\
19533747+1844596 & 9.05 & 0.02 & -27.65 & 1.05 & 7.80 & 0.16 & \\
19533964+1841466 & 12.05 & 0.02 & -26.69 & 1.21 & 6.74 & 0.30 & \\
19534282+1846129 & 11.38 & 0.05 & -24.03 & 1.37 & 7.16 & 0.42 & \\
19535983+1844498 & 11.87 & 0.02 & -24.38 & 1.11 & 6.10 & 0.21 & \\
19535325+1846471 & 8.04 & 0.01 & -23.96 & 1.05 & 8.60 & 0.17 & \\
19535764+1847570 & 10.04 & 0.02 & -27.75 & 1.26 & 6.89 & 0.18 & \\
19535064+1849075 & 9.27 & 0.03 & -27.15 & 0.77 & 7.55 & 0.17 & \\
\hline
\multicolumn{8}{l}{NGC~6791}\\
19204485+3746215 & 10.25 & 0.02 & -44.52 & 0.66 & 9.31 & 0.26 & \\
19203585+3746520 & 11.71 & 0.01 & -49.93 & 0.98 & 8.59 & 0.35 & \\
19204517+3744339 & 11.70 & 0.02 & -51.31 & 0.73 & 9.15 & 0.36 & \\
19205580+3742307 & 11.86 & 0.02 & -48.25 & 1.08 & 9.22 & 0.37 & \\
19210719+3744347 & 11.91 & 0.02 & -46.29 & 0.89 & 8.23 & 0.29 & \\
19205418+3746285 & 10.98 & 0.02 & -44.59 & 0.69 & 9.00 & 0.32 & \\
19205338+3748282 & 9.77 & 0.01 & -50.72 & 0.77 & 9.49 & 0.22 & \\
\hline
\multicolumn{8}{l}{NGC~6819}\\
19410991+4015495 & 10.08 & 0.02 & -0.12 & 0.03 & 8.04 & 0.24 & RV template star  \\
19410524+4014042 & 10.34 & 0.02 & 3.00 & 0.63 & 7.93 & 0.26 & \\
19405020+4013109 & 10.35 & 0.02 & 2.17 & 1.24 & 8.03 & 0.23 & \\
19405704+4010068 & 10.44 & 0.02 & 10.13 & 0.74 & 7.81 & 0.21 & \\
19405797+4008174 & 10.17 & 0.02 & -2.45 & 0.92 & 8.77 & 0.24 & \\
19413031+4009005 & 7.90 & 0.02 & 3.19 & 0.74 & 9.33 & 0.26 & \\
19412187+4011485 & 9.96 & 0.02 & 2.31 & 0.79 & 8.03 & 0.33 & \\
19412147+4013573 & 10.25 & 0.02 & -0.34 & 0.86 & 7.88 & 0.27 & \\
19412222+4016442 & 10.31 & 0.02 & -0.72 & 1.75 & 8.61 & 0.52 & \\
\hline
\end{tabular}
\end{center}
$^a$ Table \ref{tab-stars} includes all of the stars which were used in the analysis. 
A version of Table \ref{tab-stars} with all of the observed stars is available on the
electronic edition of MNRAS.
\end{table}

\clearpage

\begin{table}
\begin{center}
\contcaption{Sample of Cluster Giants and Derived Parameters$^a$ \label{tab-stars2}}
\begin{tabular}{lccccccccl}
\hline\hline
2MASS ID & $K_s$ & $\sigma _{K_{s}}$ & v$_r$ & $\sigma _{v_{r}}$ &
$\Sigma$W(CaT) & $\sigma _W$ & Notes \\
& (mag) & (mag) & (km s$^{-1}$) & (km s $^{-1}$)& (\AA) & (\AA) & \\
\hline
\multicolumn{8}{l}{NGC~6939}\\
20313338+6045507 & 9.80 & 0.02 & -20.31 & 0.91 & 7.69 & 0.38 & \\
20312540+6041164 & 10.03 & 0.02 & -19.95 & 0.62 & 7.45 & 0.21 & \\
20310597+6042139 & 9.56 & 0.02 & -22.01 & 0.60 & 7.75 & 0.23 & \\
20313200+6039271 & 9.81 & 0.02 & -19.80 & 0.63 & 7.93 & 0.18 & \\
20310189+6038116 & 9.72 & 0.02 & -22.47 & 0.64 & 7.63 & 0.19 & \\
20312693+6036595 & 8.55 & 0.02 & -21.22 & 0.62 & 8.33 & 0.21 & \\
20314054+6037084 & 7.97 & 0.02 & -21.90 & 0.58 & 8.40 & 0.25 & \\
20315345+6038573 & 10.23 & 0.02 & -16.76 & 0.66 & 7.56 & 0.20 & \\
20322403+6037398 & 9.80 & 0.02 & -25.87 & 0.71 & 7.59 & 0.18 & \\
20315931+6041075 & 9.95 & 0.02 & -18.77 & 0.80 & 7.45 & 0.21 & \\
20322172+6043113 & 8.33 & 0.02 & -20.93 & 0.62 & 8.44 & 0.22 & \\
20320790+6044167 & 9.83 & 0.02 & -20.76 & 0.83 & 7.66 & 0.22 & \\
20314339+6040386 & 9.84 & 0.02 & -22.99 & 0.97 & 7.77 & 0.19 & \\
\hline
\multicolumn{8}{l}{NGC~7142}\\
21450182+6549286 & 9.82 & 0.03 & -49.92 & 1.76 & 7.23 & 0.37 & \\
21444497+6549144 & 8.67 & 0.02 & -52.85 & 0.97 & 9.07 & 0.37 & \\
21443881+6546382 & 10.49 & 0.02 & -55.03 & 1.03 & 7.70 & 0.36 & \\
21450252+6545401 & 7.55 & 0.01 & -50.76 & 0.81 & 9.15 & 0.36 & \\
21454049+6544561 & 8.77 & 0.02 & -52.06 & 0.95 & 8.61 & 0.29 & \\
21452095+6547402 & 9.15 & 0.02 & -50.58 & 0.95 & 8.18 & 0.25 & \\
\hline
\multicolumn{8}{l}{NGC~7789}\\
23572445+5648304 & 7.55 & 0.02 & -55.85 & 0.87 & 8.70 & 0.27 & \\
23571692+5646200 & 8.67 & 0.01 & -55.35 & 0.76 & 8.25 & 0.24 & \\
23565473+5648163 & 9.98 & 0.02 & -53.92 & 0.14 & 7.63 & 0.22 & RV template star \\
23570324+5645580 & 5.89 & 0.01 & -55.65 & 1.12 & 9.52 & 0.28 & \\
23565546+5645091 & 8.84 & 0.02 & -54.38 & 0.18 & 8.28 & 0.23 & RV template star \\
23563303+5644332 & 10.05 & 0.02 & -59.41 & 0.04 & 7.69 & 0.23 & RV template star \\
23564587+5638407 & 9.73 & 0.01 & -57.60 & 0.79 & 7.61 & 0.23 & \\
23571400+5640586 & 8.85 & 0.01 & -57.52 & 0.83 & 8.04 & 0.24 & \\
23572501+5638363 & 9.97 & 0.02 & -59.48 & 0.73 & 8.15 & 0.38 & \\
23572872+5635228 & 10.32 & 0.02 & -55.32 & 1.13 & 6.97 & 0.28 & \\
23573184+5641221 & 8.04 & 0.01 & -59.34 & 0.95 & 8.41 & 0.26 & \\
23575156+5638566 & 5.81 & 0.02 & -55.82 & 1.03 & 9.42 & 0.27 & \\
23580133+5639219 & 9.42 & 0.02 & -59.22 & 0.82 & 7.59 & 0.27 & \\
23575204+5642256 & 6.73 & 0.02 & -53.73 & 0.89 & 9.36 & 0.27 & \\
23581624+5642054 & 10.29 & 0.02 & -59.67 & 0.89 & 8.52 & 0.25 & \\
23575501+5644323 & 8.90 & 0.02 & -56.22 & 0.86 & 8.17 & 0.22 & \\
23582319+5647371 & 8.15 & 0.01 & -56.19 & 0.85 & 8.53 & 0.24 & \\
23580015+5650125 & 8.34 & 0.02 & -54.70 & 0.87 & 8.41 & 0.23 & \\
23575149+5651040 & 8.08 & 0.01 & -56.01 & 0.92 & 8.49 & 0.27 & \\
23573079+5646443 & 10.13 & 0.02 & -55.65 & 0.75 & 7.61 & 0.22 & \\
\hline
\multicolumn{8}{l}{Be~81}\\
\multicolumn{8}{l}{\it Brighter subgroup (see \S\ref{be81-sec})}\\
19013063-0024456 & 8.58 & 0.02 & 15.84 & 0.98 & 9.61 & 0.34 & \\
19015206-0026466 & 10.61 & 0.02 & 14.24 & 1.02 & 7.81 & 0.26 & \\
19013928-0026285 & 10.19 & 0.03 & 7.82 & 1.12 & 8.27 & 0.62 & \\
\multicolumn{8}{l}{\it Fainter subgroup (see \S\ref{be81-sec})}\\
19013687-0024097 & 10.52 & 0.02 & -8.37 & 1.00 & 7.82 & 0.37 & \\
19013595-0028378 & 9.65 & 0.02 & -7.40 & 1.26 & 8.99 & 0.32 & \\
\multicolumn{8}{l}{\it Photometric outlier (see \S\ref{be81-sec})}\\
19012993-0027231 & 9.09 & 0.02 & -29.39 & 0.90 & 8.22 & 0.22 & \\
\hline
\end{tabular}
\end{center}
$^a$Table \ref{tab-stars} includes all of the stars which were used in the analysis. 
A version of Table \ref{tab-stars} with all of the observed stars is available on the
electronic edition of MNRAS.
\end{table}

\clearpage

\begin{table}
\begin{center}
\contcaption{Sample of Cluster Giants and Derived Parameters$^a$ \label{tab-stars3}}
\begin{tabular}{lccccccccl}
\hline\hline
2MASS ID & $K_s$ & $\sigma _{K_{s}}$ & v$_r$ & $\sigma _{v_{r}}$ &
$\Sigma$W(CaT) & $\sigma _W$ & Notes \\
& (mag) & (mag) & (km s$^{-1}$) & (km s $^{-1}$) &(\AA) & (\AA) & \\
\hline
\multicolumn{8}{l}{Be~99}\\
23204241+7148166 & 12.06 & 0.03 & -64.27 & 1.72 & 6.62 & 0.30 & \\
23204335+7146540 & 11.16 & 0.02 & -66.85 & 1.24 & 7.08 & 0.20 & \\
23210175+7145598 & 8.61 & 0.02 & -61.61 & 0.88 & 8.50 & 0.20 & \\
23205404+7143296 & 11.35 & 0.02 & -70.97 & 0.95 & 7.79 & 0.28 & \\
23213138+7143175 & 12.34 & 0.02 & -51.50 & 1.62 & 6.58 & 0.28 & \\
23213173+7146053 & 11.48 & 0.02 & -64.40 & 1.93 & 6.55 & 0.23 & \\
23215512+7146502 & 11.33 & 0.02 & -47.01 & 1.49 & 6.85 & 0.25 & \\
23213813+7147350 & 11.20 & 0.03 & -62.54 & 1.29 & 5.41 & 0.84 & \\
23212213+7148034 & 9.11 & 0.02 & -61.12 & 0.75 & 8.47 & 0.23 & \\
\hline
\multicolumn{8}{l}{IC~1311}\\
20104659+4112343 & 8.46 & 0.01 & -61.45 & 1.06 & 10.70 & 0.32 & \\
20103943+4114014 & 13.16 & 0.04 & -61.71 & 1.74 & 8.11 & 0.56 & \\
20104386+4110029 & 10.78 & 0.01 & -64.19 & 1.23 & 7.77 & 0.28 & \\
20103623+4107222 & 11.21 & 0.01 & -64.42 & 1.24 & 8.19 & 0.29 & \\
20105457+4112248 & 13.06 & 0.03 & -65.14 & 2.08 & 6.11 & 0.61 & \\
\hline
\multicolumn{8}{l}{King~2}\\
00504698+5815418 & 11.29 & 0.02 & -144.48 & 1.13 & 7.23 & 0.47 & \\
00503676+5808417 & 11.98 & 0.02 & -151.15 & 2.18 & 6.65 & 0.57 & \\
00505902+5808216 & 12.57 & 0.02 & -147.13 & 2.68 & 7.10 & 0.69 & \\
00511367+5807532 & 11.71 & 0.02 & -132.75 & 2.02 & 7.53 & 0.61 & \\
00510072+5810562 & 10.04 & 0.02 & -144.70 & 3.40 & 8.58 & 0.28 & \\
00511598+5813527 & 11.51 & 0.02 & -147.99 & 4.47 & 7.48 & 0.69 & \\
00505610+5812053 & 12.38 & 0.02 & -147.25 & 2.35 & 7.61 & 0.64 & \\
\hline
\multicolumn{8}{l}{NGC~7044}\\
21130270+4231244 & 10.74 & 0.02 & -49.19 & 0.98 & 8.55 & 0.27 & \\
21130119+4229295 & 11.39 & 0.02 & -55.10 & 1.53 & 6.80 & 0.47 & \\
21124797+4231110 & 11.70 & 0.02 & -50.42 & 1.60 & 8.58 & 0.55 & \\
21124114+4229229 & 10.56 & 0.01 & -52.53 & 1.15 & 8.67 & 0.46 & \\
21123464+4228075 & 10.49 & 0.02 & -49.32 & 0.93 & 8.02 & 0.29 & \\
21130646+4228414 & 10.00 & 0.02 & -49.54 & 0.93 & 8.73 & 0.35 & \\
21133247+4230482 & 10.86 & 0.02 & -53.91 & 1.09 & 8.13 & 0.34 & \\
21131398+4229449 & 10.08 & 0.01 & -49.57 & 1.15 & 8.10 & 0.31 & \\
21131533+4231270 & 11.40 & 0.02 & -50.20 & 1.04 & 8.20 & 0.30 & \\
21131452+4235234 & 11.37 & 0.02 & -49.24 & 1.10 & 7.49 & 0.34 & \\
\hline
\end{tabular}
\end{center}
$^a$Table \ref{tab-stars} includes all of the stars which were used in the analysis. 
A version of Table \ref{tab-stars} with all of the observed stars is available on the
electronic edition of MNRAS.
\end{table}

\clearpage

\begin{table}
\begin{center}
\caption{Sample of Cluster Giants from \citet{cole04} and Derived Parameters$^a$ \label{tab-stars4}}
\begin{tabular}{lccccccccl}
\hline\hline
ID$^b$ & $K_s$ & $\sigma _{K_{s}}$ & v$_r$ & $\sigma _{v_{r}}$ &
$\Sigma$W(CaT) & $\sigma _W$ \\
& (mag) & (mag) & (km s$^{-1}$) & (km s $^{-1}$) &(\AA) & (\AA) & \\
\hline
\multicolumn{8}{l}{Be~20:~\citet{mac94}}\\
022	&	14.31	&	0.08	&	74.6	&	7.5	&	6.19	&	0.03	\\
008	&	11.85	&	0.03	&	79.2	&	7.5	&	7.41	&	0.02	\\
005	&	11.35	&	0.02	&	73.9	&	7.5	&	7.56	&	0.02	\\
012	&	13.25	&	0.04	&	83.8	&	7.5	&	6.61	&	0.02	\\
\hline
\multicolumn{8}{l}{Be~39:~\citet{kal89b}}\\
KR009	&	11.43	&	0.02	&	56.7	&	7.5	&	7.14	&	0.03	\\
KR002	&	9.05	&	0.02	&	56.5	&	7.5	&	8.39	&	0.03	\\
KR012	&	11.56	&	0.02	&	55.9	&	7.4	&	7.37	&	0.03	\\
KR005	&	10.55	&	0.03	&	56.5	&	7.5	&	7.68	&	0.03	\\
KR013	&	11.56	&	0.02	&	58.2	&	7.5	&	6.87	&	0.03	\\
KR018	&	11.69	&	0.03	&	57.9	&	7.5	&	6.88	&	0.03	\\
KR017	&	11.82	&	0.02	&	54.0	&	7.6	&	6.72	&	0.03	\\
KR016	&	11.62	&	0.02	&	61.4	&	7.5	&	7.28	&	0.03	\\
KR003	&	9.79	&	0.02	&	61.0	&	7.4	&	8.26	&	0.03	\\
KR028	&	12.61	&	0.03	&	58.9	&	7.5	&	6.6	&	0.03	\\
\hline
\multicolumn{8}{l}{M67:~\citet{sand77}}\\
F104	&	8.61	&	0.02	&	33.5	&	-	&	7.13	&	0.16	\\
F164	&	7.96	&	0.02	&	33.3	&	-	&	7.4	&	0.15	\\
F105	&	7.39	&	0.02	&	34.3	&	-	&	8	&	0.16	\\
F141	&	7.94	&	0.02	&	33.6	&	-	&	7.73	&	0.14	\\
F170	&	6.49	&	0.02	&	34.3	&	-	&	8.28	&	0.19	\\
F135	&	8.95	&	0.02	&	34.3	&	-	&	7.1	&	0.13	\\
F108	&	6.49	&	0.02	&	34.7	&	-	&	8.36	&	0.17	\\
\hline
\multicolumn{8}{l}{Mel~66:~\citet{ant79}}\\
1205	&	11.82	&	0.03	&	18.3	&	7.5	&	6.86	&	0.11	\\
2261	&	10.56	&	0.02	&	41.2	&	7.5	&	7.46	&	0.16	\\
2244	&	11.83	&	0.03	&	16.4	&	7.6	&	6.97	&	0.09	\\
2233	&	12.81	&	0.03	&	14.4	&	7.4	&	6.51	&	0.09	\\
2226	&	11.20	&	0.02	&	18.9	&	7.5	&	7.11	&	0.19	\\
2133	&	9.54	&	0.02	&	10.9	&	7.5	&	7.91	&	0.12	\\
2107	&	11.77	&	0.02	&	14.9	&	7.5	&	6.57	&	0.11	\\
3260	&	11.56	&	0.04	&	16.2	&	7.4	&	6.9	&	0.12	\\
3133	&	10.96	&	0.02	&	17.2	&	7.6	&	6.98	&	0.12	\\
3235	&	11.82	&	0.02	&	18.3	&	7.7	&	6.89	&	0.10	\\
4151	&	8.84	&	0.02	&	12.7	&	7.6	&	8.23	&	0.13	\\
4265	&	10.98	&	0.02	&	7.4	&	7.5	&	6.97	&	0.12	\\
3229	&	10.93	&	0.02	&	7.7	&	7.5	&	7.08	&	0.31	\\
\hline
\end{tabular}
\end{center}
$^a$Table \ref{tab-stars4} includes all of the stars which were used in the analysis.\\
$^b$References are for star IDs. 
\end{table}

\clearpage

\begin{table}
\begin{center}
\contcaption{Sample of Cluster Giants from \citet{cole04} and Derived Parameters$^a$ \label{tab-stars5}}
\begin{tabular}{lccccccccl}
\hline\hline
ID$^b$ & $K_s$ & $\sigma _{K_{s}}$ & v$_r$ & $\sigma _{v_{r}}$ &
$\Sigma$W(CaT) & $\sigma _W$ \\
& (mag) & (mag) & (km s$^{-1}$) & (km s $^{-1}$) &(\AA) & (\AA) & \\
\hline
\multicolumn{8}{l}{NGC~104:~\citet{lee77}}\\
L5309	&	8.64	&	0.02	&	-21.8	&	7.6	&	7.98	&	0.10	\\
L5312	&	8.54	&	0.02	&	-12.3	&	7.5	&	7.91	&	0.10	\\
L5418	&	12.98	&	0.03	&	-20.6	&	7.9	&	5.77	&	0.16	\\
L5422	&	9.09	&	0.02	&	-23.1	&	7.6	&	7.38	&	0.10	\\
L5419	&	11.81	&	0.02	&	-22.2	&	7.8	&	6.13	&	0.12	\\
L5527	&	10.85	&	0.02	&	-25.5	&	7.6	&	6.79	&	0.09	\\
L5530	&	9.87	&	0.02	&	-18.6	&	7.6	&	6.92	&	0.08	\\
\hline
\multicolumn{8}{l}{NGC~1851:~\citet{stet81}}\\
003	&	10.30	&	0.02	&	324.5	&	7.4	&	7.52	&	0.02	\\
065	&	13.56	&	0.06	&	322.1	&	7.5	&	5.67	&	0.03	\\
095	&	10.09	&	0.02	&	334.1	&	7.4	&	6.94	&	0.02	\\
126	&	11.39	&	0.02	&	321.5	&	7.5	&	6.95	&	0.02	\\
123	&	14.06	&	0.06	&	329.7	&	7.4	&	4.99	&	0.03	\\
112	&	10.38	&	0.02	&	327.3	&	7.4	&	6.85	&	0.03	\\
109	&	12.12	&	0.02	&	334.3	&	7.4	&	5.96	&	0.03	\\
275	&	12.39	&	0.03	&	333.8	&	7.6	&	6.79	&	0.03	\\
160	&	13.17	&	0.04	&	326.4	&	7.7	&	5.85	&	0.03	\\
209	&	11.24	&	0.03	&	333.4	&	7.6	&	7.05	&	0.02	\\
107	&	11.52	&	0.02	&	330.7	&	7.5	&	6.78	&	0.03	\\
231	&	13.56	&	0.05	&	332.8	&	7.8	&	5.89	&	0.03	\\
175	&	13.82	&	0.05	&	332.3	&	7.9	&	6.07	&	0.04	\\
195	&	13.34	&	0.03	&	327.6	&	7.5	&	5.75	&	0.03	\\
179	&	14.23	&	0.07	&	320.4	&	7.8	&	5.66	&	0.03	\\
\hline
\multicolumn{8}{l}{NGC~1904:~\citet{stet77}}\\
6	&	12.85	&	0.03	&	203.2	&	7.4	&	4.91	&	0.03	\\
11	&	13.48	&	0.05	&	213.5	&	7.5	&	4.85	&	0.03	\\
45	&	13.21	&	0.04	&	206.8	&	7.8	&	4.95	&	0.05	\\
15	&	9.99	&	0.02	&	204.8	&	7.5	&	6.67	&	0.02	\\
241	&	10.72	&	0.02	&	228.5	&	7.5	&	5.8	&	0.02	\\
237	&	11.32	&	0.02	&	220.8	&	7.5	&	5.77	&	0.02	\\
89	&	12.13	&	0.02	&	206.5	&	7.8	&	5.51	&	0.02	\\
91	&	14.18	&	0.07	&	207.6	&	7.8	&	4.76	&	0.05	\\
224	&	13.88	&	0.05	&	220.4	&	7.9	&	4.63	&	0.05	\\
111	&	13.21	&	0.05	&	206.9	&	7.6	&	5.18	&	0.03	\\
115	&	13.72	&	0.06	&	211.0	&	7.7	&	4.65	&	0.05	\\
138	&	14.00	&	0.07	&	206.3	&	7.8	&	4.36	&	0.04	\\
153	&	10.26	&	0.02	&	206.1	&	7.4	&	6.47	&	0.02	\\
209	&	12.43	&	0.03	&	210.8	&	7.5	&	5.63	&	0.02	\\
160	&	9.61	&	0.02	&	200.3	&	7.5	&	6.42	&	0.02	\\
161	&	13.68	&	0.05	&	216.5	&	7.8	&	4.05	&	0.05	\\
176	&	12.52	&	0.03	&	215.9	&	7.6	&	4.74	&	0.02	\\
\hline
\end{tabular}
\end{center}
$^a$Table \ref{tab-stars4} includes all of the stars which were used in the analysis.\\
$^b$References are for star IDs. 
\end{table}

\clearpage

\begin{table}
\begin{center}
\contcaption{Sample of Cluster Giants from \citet{cole04} and Derived Parameters$^a$ \label{tab-stars6}}
\begin{tabular}{lccccccccl}
\hline\hline
ID$^b$ & $K_s$ & $\sigma _{K_{s}}$ & v$_r$ & $\sigma _{v_{r}}$ &
$\Sigma$W(CaT) & $\sigma _W$ \\
& (mag) & (mag) & (km s$^{-1}$) & (km s $^{-1}$) &(\AA) & (\AA) & \\
\hline
\multicolumn{8}{l}{NGC~2141:~\citet{burk72}}\\
5-09	&	11.31	&	0.02	&	33.0	&	7.5	&	7.69	&	0.03	\\
5-13	&	10.33	&	0.02	&	32.0	&	7.5	&	7.92	&	0.03	\\
4-08	&	11.58	&	0.03	&	23.2	&	7.6	&	7.76	&	0.03	\\
4-09	&	8.98	&	0.02	&	28.7	&	7.6	&	8.78	&	0.03	\\
4-13	&	11.77	&	0.03	&	32.1	&	7.4	&	6.79	&	0.03	\\
4-14	&	12.30	&	0.03	&	28.8	&	7.4	&	7.38	&	0.03	\\
3-2-52	&	10.85	&	0.02	&	33.2	&	7.5	&	7.67	&	0.03	\\
3-2-40	&	8.86	&	0.02	&	28.9	&	7.4	&	8.8	&	0.03	\\
3-2-34	&	11.57	&	0.02	&	27.6	&	7.5	&	7.26	&	0.03	\\
3-2-18	&	8.34	&	0.02	&	26.0	&	7.5	&	9.35	&	0.03	\\
1-4-05	&	10.61	&	0.02	&	50.0	&	7.5	&	7.58	&	0.03	\\
1-3-21	&	10.33	&	0.02	&	31.8	&	7.5	&	8.04	&	0.03	\\
4-25	&	10.32	&	0.02	&	31.3	&	7.5	&	8.11	&	0.03	\\
4-24	&	11.63	&	0.03	&	31.4	&	7.5	&	7.64	&	0.02	\\
\hline
\multicolumn{8}{l}{NGC~2298:~\citet{alca86}}\\
AL12	&	11.06	&	0.03	&	146.6	&	7.5	&	4.84	&	0.02	\\
AL15	&	11.77	&	0.02	&	151.2	&	7.6	&	4.31	&	0.02	\\
AL6	&	10.55	&	0.03	&	156.5	&	7.6	&	4.97	&	0.02	\\
AL22	&	12.49	&	0.03	&	162.0	&	7.4	&	4.06	&	0.02	\\
AL25	&	13.05	&	0.04	&	153.2	&	7.5	&	4.15	&	0.02	\\
\hline
\multicolumn{8}{l}{NGC~4590:~\citet{har75}}\\
I-258	&	11.86	&	0.02	&	-90.2	&	7.7	&	3.14	&	0.02	\\
I-256	&	9.81	&	0.02	&	-90.9	&	7.6	&	5.2	&	0.02	\\
I-260	&	9.51	&	0.02	&	-91.9	&	7.5	&	4.85	&	0.02	\\
I-2	&	12.83	&	0.05	&	-96.0	&	7.6	&	2.98	&	0.02	\\
I-239	&	11.71	&	0.03	&	-86.0	&	7.6	&	3.63	&	0.02	\\
I-49	&	12.68	&	0.03	&	-93.6	&	7.8	&	3.04	&	0.02	\\
I-74	&	12.20	&	0.02	&	-103.7	&	7.7	&	3.21	&	0.02	\\
I-119	&	10.92	&	0.02	&	-89.1	&	7.6	&	3.78	&	0.02	\\
II-47	&	12.76	&	0.02	&	-85.8	&	7.8	&	3.19	&	0.02	\\
\hline
\end{tabular}
\end{center}
$^a$Table \ref{tab-stars4} includes all of the stars which were used in the analysis.\\
$^b$References are for star IDs. 
\end{table}

\clearpage

\begin{table}
\begin{center}
\caption{Comparison to Previous Measurements\label{comps}}
\begin{tabular}{lcccccc}
\hline\hline
Cluster   & W$^{\prime}$ (\AA) & [Fe/H] & [Fe/H]$_{ref}^a$ & v$_r$ (km s$^{-1}$) & v$_{r,ref}$ (km s$^{-1}$) & Reference$^b$ \\
\hline
M15	&$2.17\pm0.07$&$-2.04\pm 0.07$&$-2.12\pm 0.01$&$-109.5\pm 0.0$&$-107.0\pm 0.2$&	1\\
M71	&$6.48\pm0.08$&$-0.61\pm 0.08$&$-0.70\pm 0.03$&$-25.8\pm 0.3$&$-22.8\pm	0.2$& 1	 \\
NGC 6791&$8.84\pm0.13$&$0.18\pm 0.08 $&$ 0.11\pm 0.10$&$-47.6\pm 0.3$&$-47.1\pm	0.8$& 2	 \\
NGC 6819&$8.13\pm0.11$&$-0.06\pm 0.08$&$-0.11\pm 0.06$&$-0.1\pm 0.0 $&$-7.0\pm 13.0$& 3	 \\
NGC 6939&$7.63\pm0.03$&$-0.22\pm 0.07$&$-0.19\pm 0.09$&$-21.0\pm 0.2$&$-19.0\pm	0.2$& 4	 \\
NGC 7142&$7.71\pm0.17$&$-0.20\pm 0.09$&$-0.10\pm 0.10$&$-52.0\pm 0.4$&$-44.0\pm	12.0$& 3 \\
NGC 7789&$7.61\pm0.07$&$-0.23\pm 0.07$&$-0.24\pm 0.09$&$-58.6\pm 0.0$&$-57.0\pm	7.0$& 3	 \\
\hline
\end{tabular}
\end{center}
$^a$ References to [Fe/H]$_{ref}$ values are given in Table \ref{clust1}. \\ 
$^b$ References to v$_{r,ref}$ are: (1) \citet{har96}, (2)
\citet{caro06}, (3) \citet{friel89}, and (4) \citet{mil94}.
\end{table}

\begin{table}
\begin{center}
\caption{Comparison to Previous Measurements\label{compsC04}}
\begin{tabular}{lcccc}
\hline\hline
\citet{cole04} & W$^{\prime}$ (\AA) & [Fe/H] & [Fe/H]$_{ref}^a$ & [Fe/H]$_{C04}^b$\\
Cluster&&&\\
\hline
Berkeley 20&$6.70\pm 0.03$&$-0.53\pm 0.07$&$-0.61\pm 0.14$&$-0.47\pm 0.07$\\
Berkeley 39&$7.14\pm 0.06$&$-0.39\pm 0.07$&$-0.26\pm 0.09$&$-0.32\pm 0.09$\\
M67        &$7.59\pm 0.05$&$-0.24\pm 0.07$&$-0.15\pm 0.05$&$-0.19\pm 0.05$\\
Melotte 66 &$6.82\pm 0.17$&$-0.50\pm 0.09$&$-0.47\pm 0.09$&$-0.48\pm 0.06$\\
NGC 104    &$6.17\pm 0.06$&$-0.71\pm 0.07$&$-0.70\pm 0.07$&$-0.66\pm 0.09$\\
NGC 1851   &$5.41\pm 0.09$&$-0.96\pm 0.08$&$-0.98\pm 0.06$&$-0.96\pm 0.12$\\
NGC 1904   &$4.55\pm 0.07$&$-1.25\pm 0.07$&$-1.37\pm 0.01$&$-1.37\pm 0.11$\\
NGC 2141   &$7.48\pm 0.07$&$-0.27\pm 0.07$&$-0.33\pm 0.10$&$-0.26\pm 0.10$\\
NGC 2298   &$3.07\pm 0.08$&$-1.74\pm 0.07$&$-1.74\pm 0.06$&$-1.69\pm 0.07$\\
NGC 4590   &$2.27\pm 0.10$&$-2.00\pm 0.08$&$-1.99\pm 0.10$&$-2.07\pm 0.09$\\
\hline
\end{tabular}
\end{center}
$^a$ References to [Fe/H]$_{ref}$ values are given in Table \ref{clust1}.\\
$^b$ Metallicity values defined by equation 5 of \citet{cole04}.
\end{table}

\clearpage

\begin{table}
\begin{center}
\caption{Comparison to Previous Measurements\label{dmclust}}
\begin{tabular}{lcccc}
\hline\hline
Cluster& [Fe/H]$_{ref}$ & [Fe/H]$_{line}$ &[Fe/H]$_{quad}$ &Reference$^a$\\
\hline
Berkeley 20&-0.44 $\pm$ 0.13&-0.43 $\pm$ 0.13&-0.54 $\pm$ 0.10& 8\\
M67        &-0.03 $\pm$ 0.03&-0.08 $\pm$ 0.13&-0.10 $\pm$ 0.10& 7\\
Melotte 66 &-0.38 $\pm$ 0.15&-0.38 $\pm$ 0.15&-0.48 $\pm$ 0.13& 5\\
NGC 104    &-0.67 $\pm$ 0.03&-0.63 $\pm$ 0.13&-0.77 $\pm$ 0.10& 2\\
NGC 1851   &-1.27 $\pm$ 0.03&-0.93 $\pm$ 0.13&-1.09 $\pm$ 0.11& 9\\
NGC 1904   &-1.37 $\pm$ 0.05&-1.26 $\pm$ 0.13&-1.41 $\pm$ 0.10& 3\\
NGC 2141   &-0.18 $\pm$ 0.15&-0.12 $\pm$ 0.13&-0.16 $\pm$ 0.11& 8\\
NGC 2298   &-1.74 $\pm$ 0.06&-1.84 $\pm$ 0.13&-1.88 $\pm$ 0.10& 3\\
NGC 4590   &-2.00 $\pm$ 0.03&-2.15 $\pm$ 0.14&-2.09 $\pm$ 0.10& 3\\
M15        &-2.12 $\pm$ 0.01&-2.19 $\pm$ 0.13&-2.11 $\pm$ 0.10& 3\\
M71        &-0.70 $\pm$ 0.03&-0.51 $\pm$ 0.13&-0.63 $\pm$ 0.11& 3\\
NGC 6791    &0.47 $\pm$ 0.04& 0.41 $\pm$ 0.14& 0.56 $\pm$ 0.12& 4\\
NGC 6819    &0.09 $\pm$ 0.03& 0.13 $\pm$ 0.14& 0.18 $\pm$ 0.12& 1\\
NGC 6939    &0.06 $\pm$ 0.06&-0.06 $\pm$ 0.13&-0.08 $\pm$ 0.10& 6\\
NGC 7142    &0.14 $\pm$ 0.06&-0.03 $\pm$ 0.15&-0.04 $\pm$ 0.13& 6\\
NGC 7789   &-0.04 $\pm$ 0.05&-0.07 $\pm$ 0.13&-0.09 $\pm$ 0.11& 7\\
\hline
\end{tabular}
\end{center}
Predicted metallicities when high-dispersion metallicity estimates from multiple
sources are used in the calibration.\\
$^a$ References to [Fe/H]$_{ref}$ values are: (1) \citet{brag01}, (2) \citet{car07b}, 
(3) \citet{cg97}, (4) \citet{grat06}, (5) \citet{grat94}, (6) \citet{jac08}, 
(7) \citet{tat05}, (8) \citet{yon05}, (9) \citet{yon08}.
\end{table}

\begin{table}
\begin{center}
\caption{Cluster Velocities and Metallicites \label{tab-clus}}
\begin{tabular}{lccccc}
\hline\hline
Cluster  & W$^{\prime}$ (\AA) & [Fe/H]& v$_r$ (km s$^{-1}$)            \\
\hline
Be~81   &7.85 $\pm$ 0.25&$-$0.15 $\pm$ 0.11&    13.00 $\pm$	4.24	\\
Be~99   &6.57 $\pm$ 0.23&$-$0.58 $\pm$ 0.10& $-$62.28 $\pm$	7.46 	\\
IC~1311 &7.40 $\pm$ 0.44&$-$0.30 $\pm$ 0.16& $-$63.15 $\pm$	1.79 	\\
King~2	&7.04 $\pm$ 0.15&$-$0.42 $\pm$ 0.09&$-$144.25 $\pm$	5.92 	\\
NGC~7044&7.83 $\pm$ 0.17&$-$0.16 $\pm$ 0.09& $-$50.56 $\pm$	2.18 	\\
\hline
\end{tabular}
\end{center}
\end{table}

\clearpage

\clearpage

\clearpage

\clearpage

\clearpage

\clearpage

\clearpage

\clearpage

\clearpage

\clearpage

\clearpage

\clearpage

\clearpage

\clearpage

\end{document}